# Titan's atmosphere as observed by Cassini/VIMS solar occultations: $CH_4$, CO and evidence for $C_2H_6$ absorption

Luca Maltagliati[a,*], Bruno Bézard[a], Sandrine Vinatier[a], Matthew M. Hedman[b], Emmanuel Lellouch[a], Philip D. Nicholson[c], Christophe Sotin[d], Remco J. de Kok[e], Bruno Sicardy[a]

[a] LESIA-Observatoire de Paris, CNRS, Univ. Paris 6, Univ. Paris 7, 5 place Jules Janssen, 92195 Meudon Cedex, France
[b] Department of Physics, University of Idaho, Moscow, ID 83844, United States
[c] Department of Astronomy, Cornell University, Ithaca, NY 14853, United States
[d] Jet Propulsion Laboratory, California Institute of Technology, 4800 Oak Grove Drive, Pasadena, CA 91109, United States
[e] SRON, Sorbonnelaan 2, 3584 CA Utrecht, Netherlands

*Abstract*

We present an analysis of the VIMS solar occultations dataset, which allows us to extract vertically resolved information on the characteristics of Titan's atmosphere between ~100–700 km with a vertical resolution of ~10 km. After a series of data treatment procedures to correct problems in pointing stability and parasitic light, 4 occultations out of 10 are retained. This sample covers different seasons and latitudes of Titan. The transmittances show clearly the evolution of the haze, with the detection of the detached layer at ~310 km in September 2011 at mid-northern latitudes. Through the inversion of the transmission spectra with a line-by-line radiative transfer code we retrieve the vertical distribution of $CH_4$ and CO mixing ratio. For methane inversion we use its 1.4, 1.7 and 2.3 μm bands. The first two bands are always in good agreement and yield an average stratospheric abundance of 1.28±0.08%, after correcting for forward-scattering effects, with no significant differences between the occultations. This is significantly less than the value of 1.48% obtained by the GCMS/Huygens instrument. We find that the 2.3 μm band cannot be used for the extraction of methane abundance because it is blended with other absorptions, not included in our atmospheric model. The analysis of the residual spectra after the inversion shows that such additional absorptions are present through a great part of the VIMS wavelength range. We attribute many of these bands, including the one at 2.3 µm, to gaseous ethane, whose near-infrared spectrum is not well modeled yet. Ethane also contributes significantly to the strong absorption at 3.2–3.5 µm that was previously attributed only to C–H stretching bands from aerosols. Ethane bands may affect the surface windows too, especially at 2.7 µm. Other residual bands are generated by stretching modes of C–H, C–C and C–N bonds. In addition to the C–H stretch from aliphatic hydrocarbons at 3.4 µm, we detect a strong and narrow absorption at 3.28 µm which we tentatively attribute to the presence of PAHs in the stratosphere. C–C and C–N stretching bands are possibly present between 4.3–4.5 µm. Finally, we obtain the CO mixing ratio between 70 and 170 km, through the inversion of its 4.7 µm band. The average result of 46±16 ppm is in good agreement with previous studies.

---

[*] Corresponding author. Address: CEA-Saclay, DSM/IRFU/Service d'Astrophysique, Centre de l'Orme des Merisiers, bât. 709, 91191 Gif/Yvette Cedex France. Email address: luca.maltagliati@obspm.fr and luca.maltagliati@cea.fr.

*1. Introduction*

The Cassini/Huygens spacecraft has increased significantly our characterization of the vertical structure of Titan's atmosphere. Several components were measured directly in-situ during the descent of the Huygens probe: temperature (Fulchignoni et al. 2005), radiative budget (Tomasko et al. 2008a), aerosol physical parameters (Tomasko et al. 2008b), and the abundance of methane and other constituents (Niemann et al. 2005, 2010) in the lower atmosphere. After Huygens, the Composite InfraRed Spectrometer (CIRS) onboard Cassini is monitoring routinely Titan's vertical atmosphere to extract latitude/pressure profiles of temperature (Achterberg et al. 2011), gaseous constituents (e.g. Teanby et al. 2009a,b, Vinatier et al. 2010a) and aerosol (Vinatier et al. 2010b, Anderson & Samuelson 2011). These results, corroborated by the Imaging Science Subsystem (ISS) images (e.g. West et al. 2011), give important constraints on the global dynamics of Titan's atmosphere and its seasonal and latitudinal variations.

A powerful technique for the observation of atmospheric vertical profiles, used often on Earth, is the occultation measurement. During an occultation an instrument observes a source of light – the Sun, a star, or a spacecraft radio signal – both outside and inside the atmosphere, at different altitudes over the planet. By dividing the spectra obtained inside the atmosphere with a reference spectrum of the source of light outside the atmosphere one can recover the atmospheric spectral signatures. Occultations are self-calibrating: a simple ratioing allows us to eliminate potential systematic effects, provided the instrument is stable in the course of an occultation. Solar and stellar occultations from spacecraft have been successfully used to study the atmospheres of Mars (e.g. Forget et al., 2009, Maltagliati et al. 2013) and Venus (e.g. Royer et al. 2010, Belyaev et al. 2012).

Cassini's payload includes two instruments that can perform occultations: the UltraViolet Imaging Spectrograph (UVIS) and the Visual and Infrared Mapping Spectrometer (VIMS). UVIS is sensitive to the upper atmosphere between 500–1400 km (Koskinen et al. 2011), a very important region to study the sources and sinks of the various molecules leading to aerosol formation (Lavvas et al. 2011). In this paper we will analyze solar occultations by VIMS. Results from the first VIMS solar occultation (January 2006) were presented in Bellucci et al. (2009). They extracted the $CH_4$ profile between 100 and 700 km from the 2.3 µm band and the average CO volume mixing ratio, from its 4.7 µm band, between 70 and 130 km. They also analyzed the aerosol extinction vertical profiles from the methane windows and inferred the haze number density and dimension of the fractal aggregates forming the aerosol as a function of altitude. An important result was their detection of a strong absorption band centered at 3.4 µm, which they identified as C–H stretching transitions from the haze particles. While their results are in general agreement with previous studies, they left some open questions. In particular, they found an increase in methane abundance below 250 km, up to 2.2% at 100 km, in disagreement with in-situ measurements (Niemann et al. 2010) and with our knowledge of Titan's dynamical and chemical processes in the lower atmosphere that predict a constant $CH_4$ abundance above 30 km.

In this paper we present an analysis of the whole VIMS solar occultations dataset available up to March 2014, which spans different seasons and latitudes of Titan. We focus in particular on the extraction and interpretation of the vertical profiles of two gaseous components, $CH_4$ and CO, using a line-by-line radiative transfer code that includes up-to-date libraries of spectral lines of the most abundant gaseous molecules in Titan's atmosphere. Finally, we allot special attention to the detection and identification of additional absorptions that appear in VIMS transmission spectra, including within the 2.3 µm methane band previously mentioned.

## 2. Dataset description

### 2.1 The VIMS/Cassini solar occultations

VIMS is an imaging spectrometer mounted onboard the Cassini spacecraft (Brown et al. 2004). It is composed of two main channels, covering respectively the 0.3–1.05 µm (called visible channel) and the 0.89–5.1 µm (infrared channel) wavelength ranges. The spectral range is covered by 352 spectral pixels (*spectels*), 96 of which dedicated to the visible channel and the other 256 to the infrared one. During solar occultations only the infrared channel is active. The FWHM of the spectels varies through the full wavelength range, progressively increasing towards longer wavelengths, from ~13 nm to ~20 nm.

VIMS employs various modes of observations, in nadir, limb and occultation geometry. It performs solar occultations using a dedicated optical port, inclined by 20° with respect to the main boresight. This optical configuration has been devised because sunlight is much stronger than any other signal from Titan and it needs to be attenuated before reaching the detector, which is shared with the main port. The attenuation factor is $\sim 2.5 \cdot 10^7$. Occultations can be in ingress (the Sun is progressively occulted by the planet) or in egress (the spacecraft comes progressively out from the planet's shadow). For the rest of the paper, ingress occultations are indicated with *I* after the flyby number and the egress ones with *E* (e.g. "T10I"). This geometrical difference does not have any other impact on the measurements: egress and ingress occultations are treated in the same way.

The VIMS-IR channel is a 1x256 pixels linear array (one pixel for each wavelength), so that each observed spatial coordinate is acquired by the same detector. The spatial mapping, up to 64x64 pixels, is obtained by a scanning mirror. The field-of-view (FOV) for an observation is set in advance. For the first four occultations, a 12x12 pixels FOV was chosen. Subsequently the FOV was reduced to 8x8 pixels (see Table 1 for details). The angular resolution is 0.5x0.5 mrad per pixel. The image of the Sun on the VIMS detector extends over approximately 2x2 pixels, so we do not lose any signal from the Sun when it is within the FOV (Fig. 1). The exposure time for each image is 20 or 40 ms per pixel depending on the occultation (Table 1). In solar occultations we do not exploit the imaging capability of VIMS and we extract one spectrum per image, following a procedure described in detail in Sect. 3.1. In this way we obtain a full spectrum of the infrared channel for each probed altitude. A full occultation lasts ~15 minutes on average.

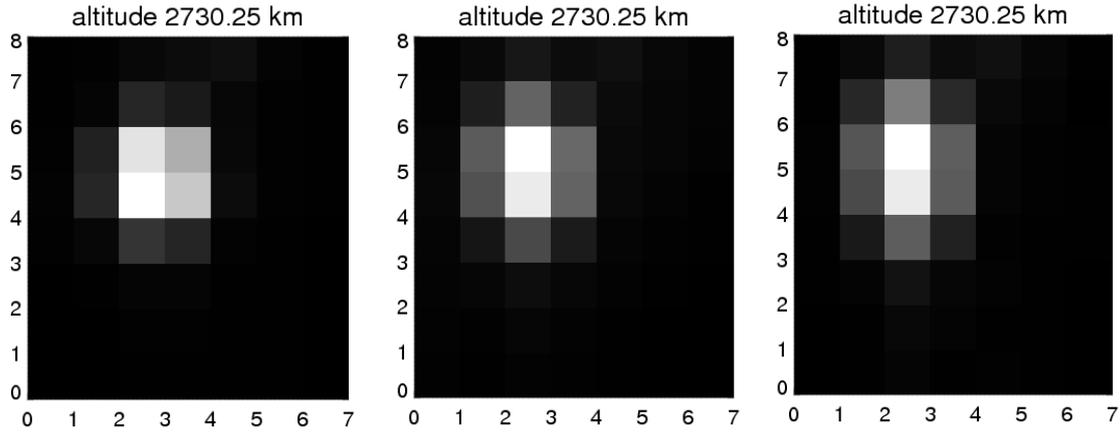

Figure 1: Example of Sun imaging by VIMS outside the atmosphere at different wavelengths from the T78E occultation. Left: 1 μm. Middle: 2 μm. Right: 2.8 μm. Slight variations of the Sun's image with wavelength can be noticed.

As the spacecraft moves into (or out of) Titan's shadow, VIMS is able to observe light passing through different altitudes. The vertical resolution varies from 5 to 15 km depending on the occultation (except one with a resolution of 40 km, which is not used in the present work). This allows the detection of much smaller vertical variations than the typical scale height of Titan's stratosphere, ~40 km.

The reconstruction of observational geometry is crucial for occultations. In particular, the coordinates of the surface point closest to the optical path, as well as the altitude of each spectrum over the surface (also called *impact parameter*) must be known with precision. In order to compute this information we built dedicated procedures employing the SPICE system from NASA's Navigation and Ancillary Information Facility (NAIF). SPICE accounts for a purely geometric reconstruction of the sunrays, and does not take into consideration refraction effects. Tests for VIMS occultation observations show that atmospheric refraction is negligible at the altitudes considered in the present study (Bellucci et al. 2009). All VIMS occultations start acquiring data well above the first detection of atmospheric signatures, which occurs at ~700 km. The highest observed altitude varies between 1300 and 7000 km, depending on the occultation. This provides a reference solar spectrum for each occultation, an important condition for the correct removal of instrumental effects.

Table 1 summarizes the main characteristics of each occultation. VIMS has acquired 10 solar occultations since the beginning of the mission, spanning different seasons with a special focus on the northern spring equinox period, and with good latitudinal sampling. Data from flyby T58 had to be immediately discarded because it could not image the Sun at all, due to a pointing error. Other two occultations have been acquired during the T103 flyby (July 2014) but they will not be analyzed here.

Table 1: main characteristics of all solar occultations acquired by VIMS between January 2006 and September 2011.

| *flyby* | *month* | *geometry* | *Approx. latitude (°N)* | *S/C-Titan distance (km)* | *Average vertical resolution (km)* | *Exposure time (ms)* | *Image dimensions (pixels)* |
|---|---|---|---|---|---|---|---|
| T10 | Jan. 2006 | Ingress | -40 | 5800 | 15 | 40 | 12x12 |
| T10 | Jan. 2006 | Egress | -70 | 8300 | 15 | 40 | 12x12 |
| T26 | Jan. 2007 | Ingress | -76 | 5700 | 40 | 20 | 12x12 |
| T32 | Apr. 2007 | Ingress | -45 | 16100 | 15 | 40 | 12x12 |
| T53 | Apr. 2009 | Egress | +1 | 6300 | 7 | 20 | 8x8 |
| T58 | Jul. 2009 | Ingress | +81 | 10400 | --- | 20 | 8x8 |
| T62 | Oct. 2009 | Ingress | -17 | 20000 | 5 | 40 | 8x8 |
| T62 | Oct. 2009 | Egress | -76 | 7300 | 4 | 40 | 8x8 |
| T78 | Sept. 2011 | Ingress | +40 | 9700 | 10 | 20 | 8x8 |
| T78 | Sept. 2011 | Egress | +27 | 8400 | 10 | 20 | 8x8 |

## 2.2 Occultations' overview and present issues

A good occultation needs to satisfy two basic prerequisites. First, the intensity variation on the detector (for each wavelength) during an occultation must be due, within the noise, only to atmospheric extinction. Only in this case we can remove all spurious contributions, of instrumental or external origin, by ratioing each spectrum by the reference solar spectrum. Second, the pointing must be stable, otherwise intensity fluctuations can occur with the movement of the Sun in the FOV, because of the different pixels (or intra-pixel) response, or in more extreme cases because the Sun moves out of the detector. If these conditions are satisfied, the lightcurve for each spectel is constant until atmospheric absorption, due to gases and/or aerosols, attenuates the signal progressively to zero.

Figures 2 and 3 show respectively the lightcurves of the total intensity on the FOV with respect to altitude (time) in the 1 μm window and the position of the Sun's center of all occultations (except T58). The same behavior is seen at all wavelengths.

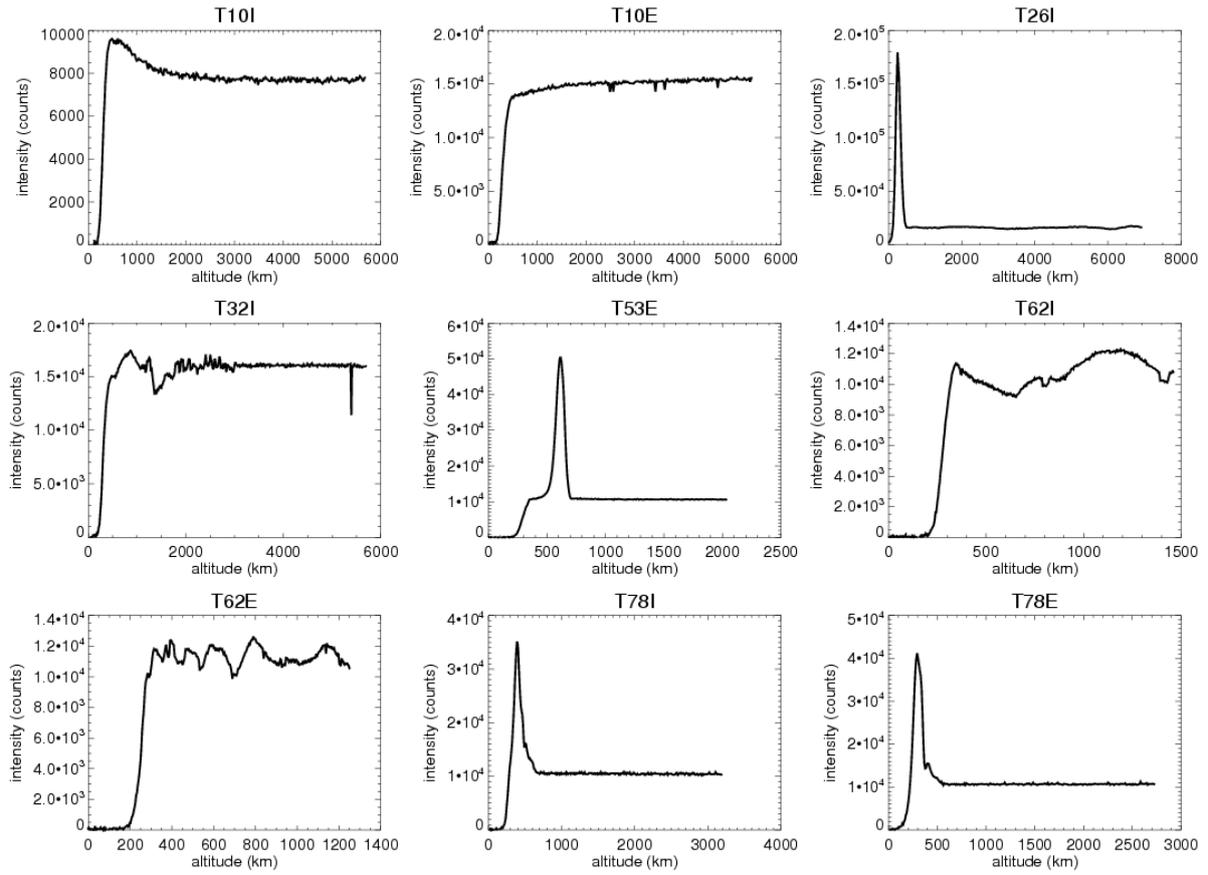

**Figure 2: Total intensity on the FOV as a function of altitude at 1μm for the nine occultations that imaged the Sun.**

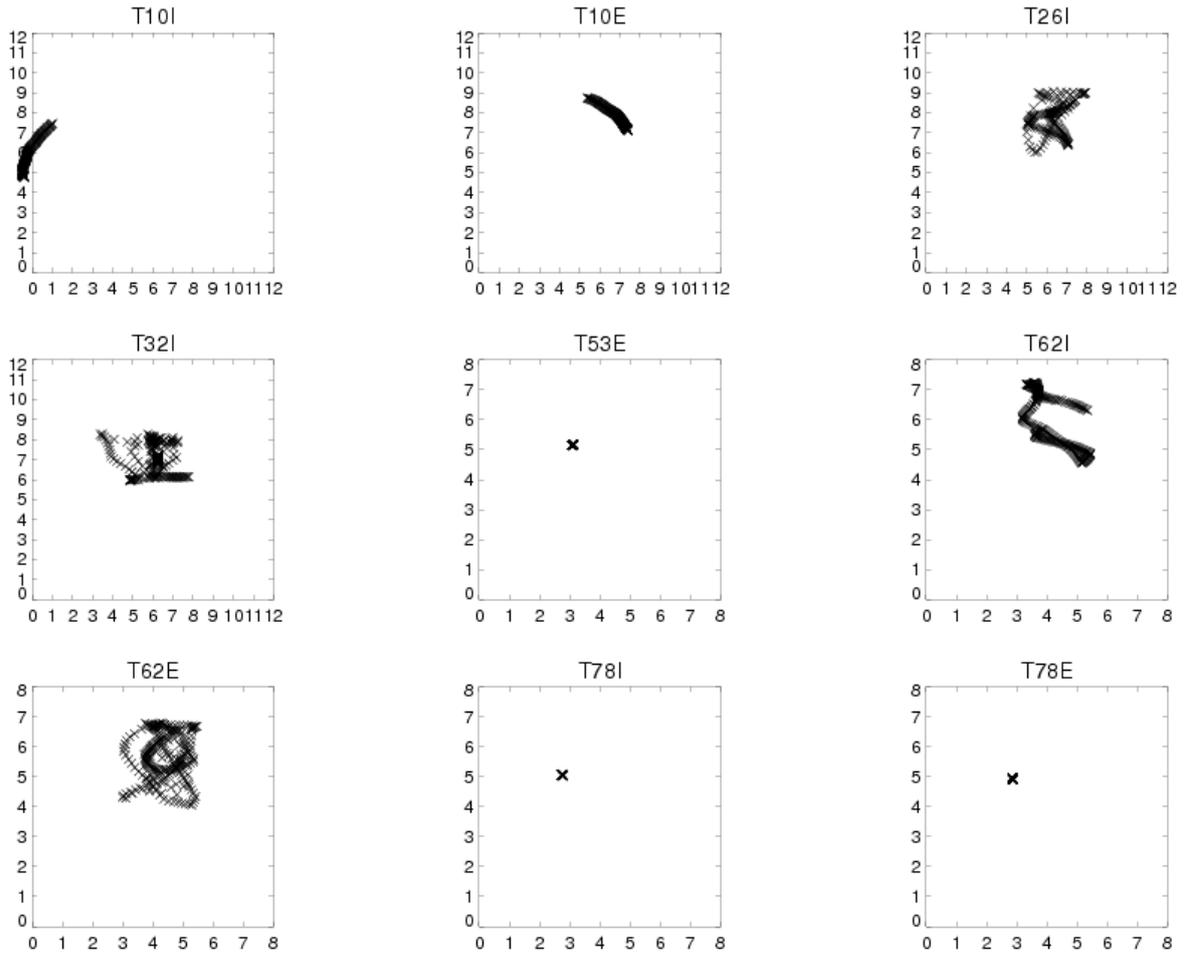

Figure 3: Evolution of the coordinates of the Sun's center in the detector for all VIMS solar occultations that imaged the Sun. Each cross corresponds to an altitude. In the case of T10I the Sun is partially out of the FOV at the beginning of the occultation and the reconstruction of the position of the center is approximate. The wide variations observed for T36I, T32I, T62I and T62E mark the activation of thrusters during the occultation. For T53E, T78I and T78E the crosses for the various altitudes are largely superimposed on each other, indicating the stability of the pointing.

All the lightcurves of Fig. 2 present some issues. Looking at the Sun's position helps to understand which issue affects a specific occultation. Figure 3 shows that for T10I the Sun is only partially within the detector at the beginning, but starts entering the FOV at ~1500 km, creating the observed flux increase. In four cases, orbits T26I, T32I, T62I and T62E, Cassini activated the thrusters during the occultation, inducing a significant pointing instability and consequently a strong fluctuation in the intensity. In the case of orbit T32I this happens in the middle of the occultation, at a tangent altitude of ~3000 km, as it can be seen by the change from a stable regime to a shaky one. Four other occultations – T26I, T53E and both T78 – exhibit a strong spike within the atmosphere, with a maximum at ~500 km for orbit T53E and ~300 km for the other three. The increase in intensity is abrupt and accounts for a factor of several units with respect to the normal level. The T53E observation shows that the falling back to pre-spike values is also very rapid and happens within a few tens of kilometers. In the case of the other three orbits the analysis of the decrease is more difficult because it is intertwined with the effect of aerosol extinction, but it seems to have the same behavior as orbit T53E. This spike is likely connected to an individual event during the occultation and not to pointing problems, because the T53E and the two T78 observations are very stable (Fig. 3). The abrupt intensity increase is most probably caused by a specific geometric configuration of VIMS' solar occultations, linked to the fact that the main boresight is not shuttered during an occultation. In the majority of the observations and for most of the time, the main port points at the background sky which does not influence the detected flux. It can happen however that, in the course of an occultation, the main port observes the illuminated limb of Titan. In this case the light from the limb collected by the main boresight hits the detector. Because this stray light is not attenuated as the solar port is, its contribution to the signal is higher than the signal from the solar port itself. This intensity spike ends when the main boresight stops receiving light from Titan's limb. An inspection of the images obtained during this contamination highlights an additional problem: this spurious signal does not just induce a simple constant increase of the background level, but has a spatially variable shape on the FOV that evolves significantly and rapidly with time (Fig. 4). The remaining occultation, T10E, does not exhibit any of these strong spurious features, but a long-term drift can be noticed in the form of a slow intensity decrease even outside the atmosphere. This effect, already observed by Bellucci et al. (2009), is probably due to the slow movement of the Sun within the FOV (see Fig. 3) and the consequent variation of the detector's response.

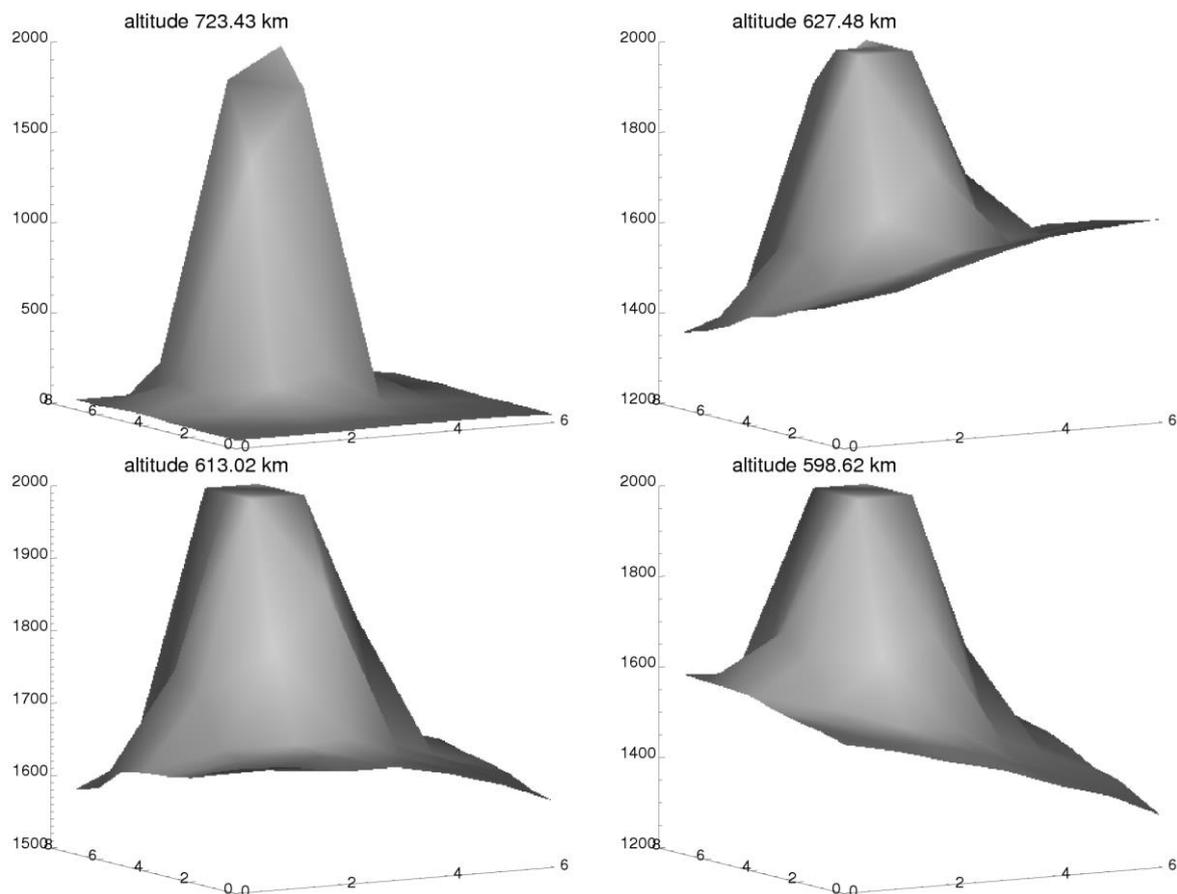

**Figure 4:** Example of the fast and unpredictable evolution of the spurious signal from the main boresight on the FOV for occultation T53E. The x- and y-axis indicate the pixel number and the z-axis the intensity in counts units. Note that the range of the z-axis changes between the figures. The top left plot shows a non-contaminated image for comparison. The other three images exhibit a reversal in the direction of the parasitic light in less than 30 km. The strong difference in background intensity with respect to the non-contaminated image can be noticed. The intensity counts are limited to 2000 for visual clarity.

## 3 Data Treatment

### 3.1 From image intensity to atmospheric transmittance

Due to the issues presented in the previous paragraph, the VIMS solar occultations need pre-treatment to extract a useable signal. A dedicated procedure was built to separate the Sun's intensity of each spatial image (at each wavelength and altitude) from the various spurious contributions detailed in the previous Section. We assumed that the Sun could be represented with a bi-dimensional Gaussian function and the background – including all the contaminations – with a second-order surface. A second-order surface simulates the contamination of light from the main boresight better than a constant background or a simple plane. If this contamination is not present, a constant level of background suffices.

An iterative procedure is set for each spatial image. First the image is fitted with an elliptical Gaussian surface using the IDL routine *mpfit2dpeak*. Seven parameters are used in the fit: a constant background level, the peak intensity and its coordinates on the image, the two FWHM for the two axes of the Gaussian, and the tilt angle with respect to the axes of the image. Since we do not expect the apparent size of the Sun at a given wavelength to vary with altitude (we can neglect diffraction-induced distortions of the solar image), we hold the two FHWM fixed at the value found at the highest altitude. After the Gaussian fit, a second-order surface is fitted to the image of the residuals. The result of this fit is subtracted from the original image and another Gaussian fit is performed on the resulting image. The full procedure is repeated once. At the end, the majority of the background variations are eliminated and only the different response of the pixels of the image remains. Figures 5 and 6 illustrate a case without and with boresight contamination respectively. The corrected images, as well as the images of the final residuals, are very similar, both in shape and intensity.

The final intensity, for each wavelength and altitude, is the integral of counts over the final corrected image. The contribution of the residuals to the final intensity is negligible. We also tried to fit simultaneously the Gaussian and the second-order surface to the image to extract the intensity of the Sun and the residuals' correction all at once, with an ad-hoc modification of the *mpfit2dpeak* routine, but this 12-parameters fit sometimes had convergence problems and we stuck to our first approach.

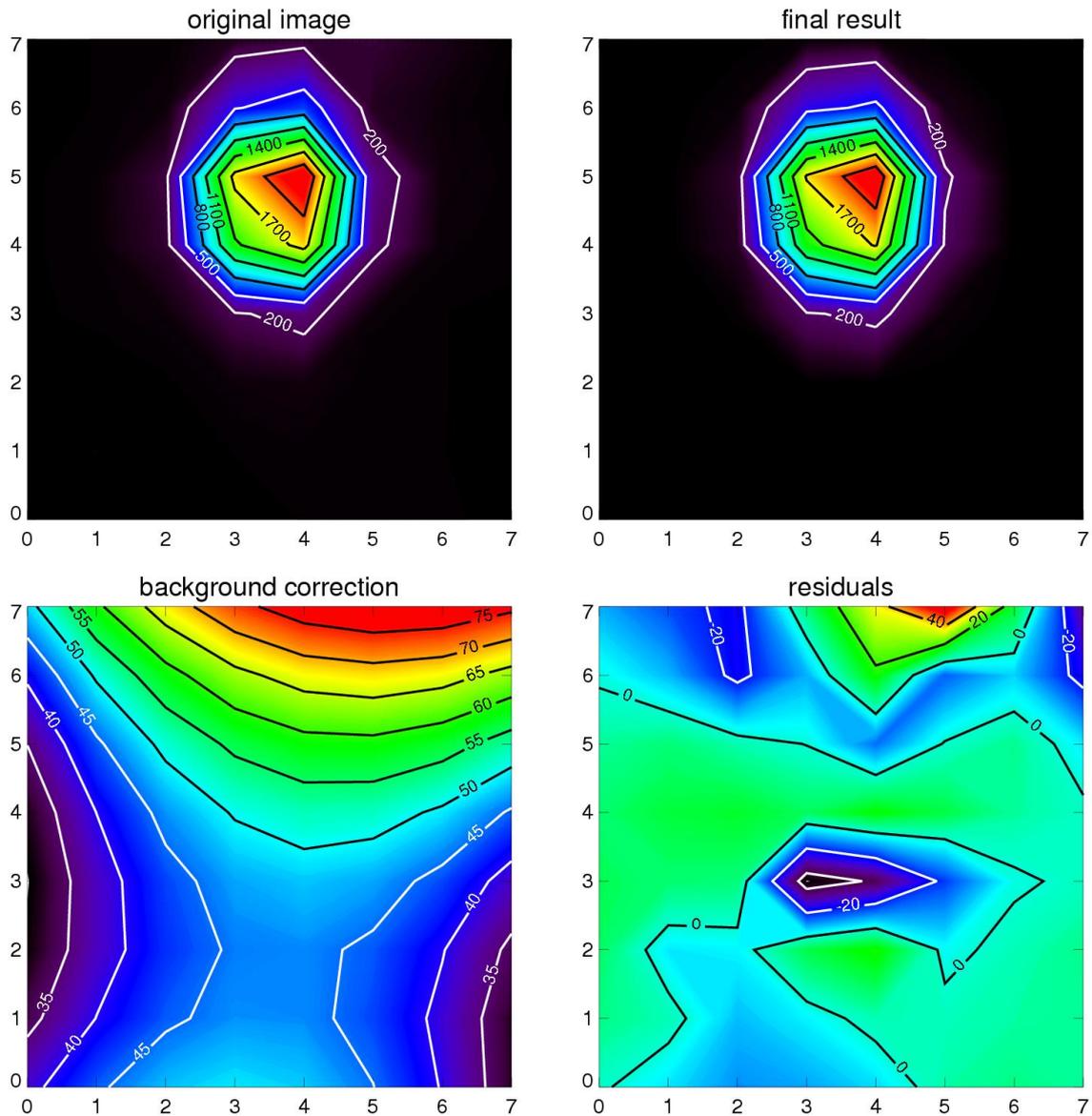

**Figure 5: Example of the data treatment procedure. This is the 1 μm image at 1448.6 km of the T53E occultation, typical of a case without stray light. Top left: the starting untreated image. Top right: the final product. Bottom left: the surface used to correct the background. Bottom right: the final residual image. In each image the color, together with the contour levels, indicates the intensity counts for each panel. The color scale changes between images. The contours are regularly spaced in each image (they span between 200 and 2000 with a step of 300 in the two top images; from 35 to 75 with a step of 5 in the background correction; from -40 to +40 with a step of 20 in the final residuals).**

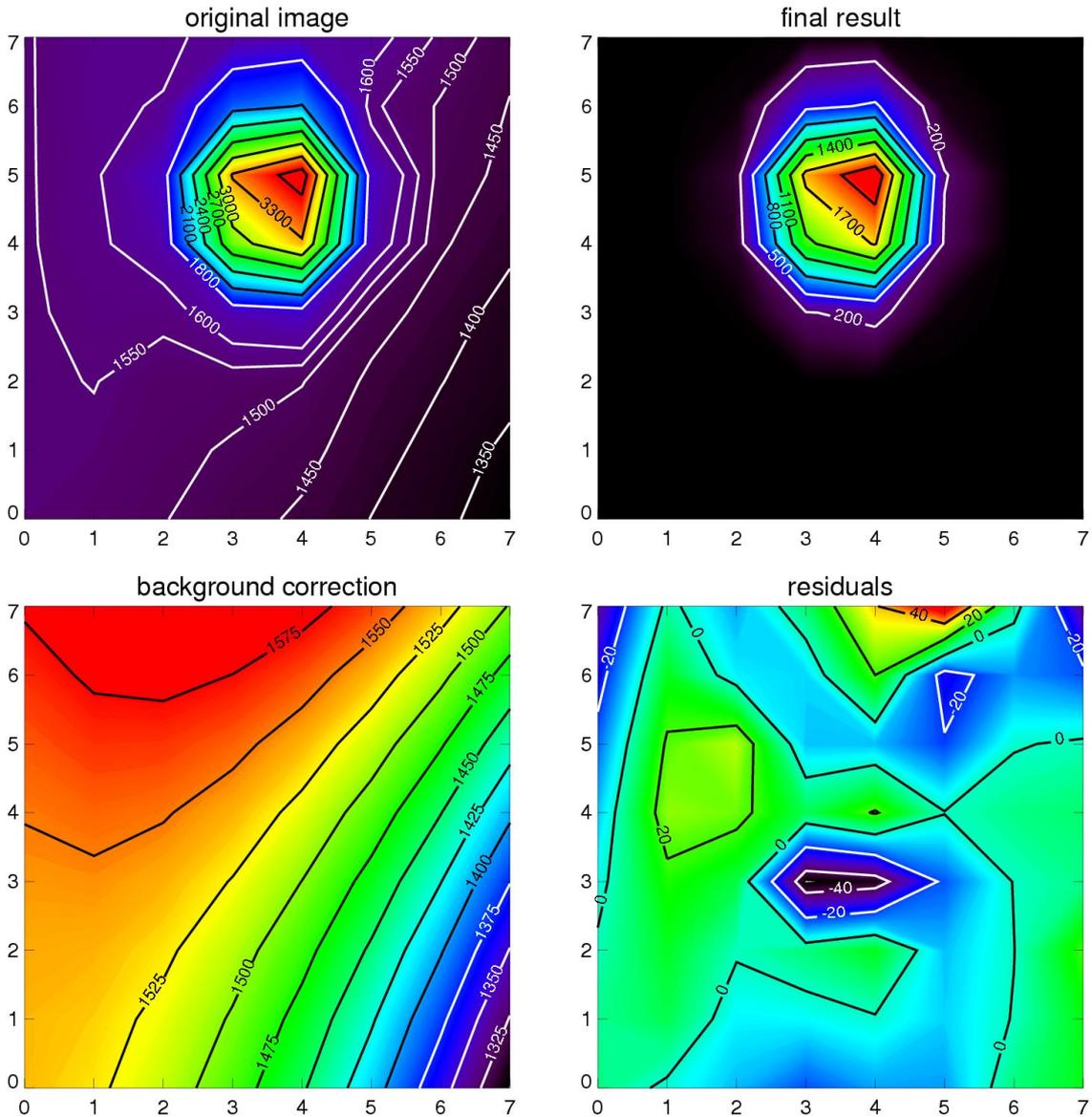

**Figure 6:** Same as Fig. 5, but for an altitude where the impact of the parasitic light from the main boresight is strong (598.6 km). The original image shows the effect of the stray light, both on the shape of the contours and on the counts level (that starts at 1350 instead of 200 as in Fig. 5). The correction surface (bottom left) changes consequently. The images of the final result and of the residuals (right plots) are color-scaled as the corresponding ones of Fig. 5 to underline the similarity between the products of the correction procedures in the two cases.

Not all orbits could be corrected with this procedure. We were not able to remove the fluctuations due to thrusters' activation, because the rapid motion of the solar image does not permit to obtain a consistent fit. The four affected orbits (T26I, T32I, T62I and T62E) were thus excluded from the analysis. The T10I occultation was also eliminated, because we could not reconstruct the flux variation caused by the progressive entrance of the Sun in the image.

At the end we have to restrict our analysis to four occultations: T10E, T53E, T78I and T78E. The T10E has also been analyzed by Bellucci et al. (2009) and it is interesting to compare the results by using different image treatment methods and a different radiative transfer code (see Sect. 3.2). In the case of the stable occultations (T53E and both T78) we found that we could obtain a smoother lightcurve for each wavelength by cutting a sub-image of 5x5 pixels around the Sun's center.

We extract the transmittance from the intensity by dividing each spectrum within an occultation by a reference solar spectrum (Fig. 7). Here by "solar spectrum" we mean the solar spectrum as observed by VIMS outside Titan's atmosphere, i.e. convolved with the spectral response of the instrument. We create a dedicated reference solar spectrum for each occultation, by averaging all spectra acquired between 1000 and 2000 km. Note that the T10E occultation exhibits also five isolated altitudes, outside of the atmosphere, where there is a sudden decrease of the intensity. These are identified by the spikes in its Fig. 2 plot. These spikes have been removed by linear interpolation between the four closest correct values around them (two per side) for each wavelength.

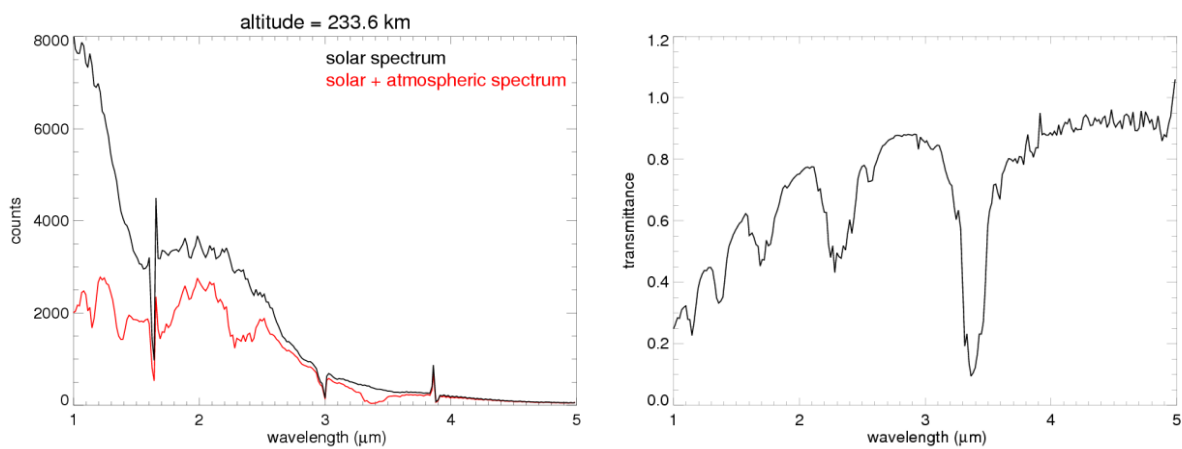

**Figure 7: Left: the "solar" reference spectrum (black) and the spectrum measured at 233.6 km (red) for occultation T78E. In addition to the atmospheric absorptions, bad spectels can be seen around 1.65 μm, 2.90 μm and 3.85 μm. These spikes happen at the locations where the order-sorting filters of VIMS are located (Brown et al. 2004). Right: the resulting transmittance after ratioing the red spectrum to the black one.**

The transmittance outside the atmosphere at each wavelength is then fitted with a polynomial. This step is necessary to take into account the long-term drifts as observed most evidently in the case of T10E (Fig. 2). For this latter occultation we also tried different fitting functions but none could manage to reproduce the shape of the drift, particularly below ~1300 km, as well as a polynomial. The order of the polynomial is evaluated empirically for each occultation and is carefully selected in order to remove the drift without introducing additional spurious effects. We employ a $4^{th}$-order polynomial for the T10E and a simple linear trend for the other orbits (Fig. 8). The correction is extrapolated at lower altitudes. The spectra change accordingly (Fig. 9). Figure 10 shows an example of comparison between the original intensity and the final transmittance for all occultations at one wavelength (2 µm). Most parasitic effects have been removed. Minor residuals still persist, as for example the slight bump visible in the transmittance of occultation T53E at ~500–550 km. The unusual slope at the beginning of the atmospheric extinction for both T78 occultations, at ~300 km, can also be due to an imperfect removal of the stray light, considering that for these occultations the boresight effect coincides with the onset of the aerosol extinction. However, this slope happens at altitudes where we detect the signature of a detached layer (see below) so it might be a real effect linked to a change in behavior or density of aerosols. Because the inversion method is very sensitive even to small transmittance variations, these residuals have a non-negligible effect on the inferred vertical profiles of the gases (see Sect. 4.1).

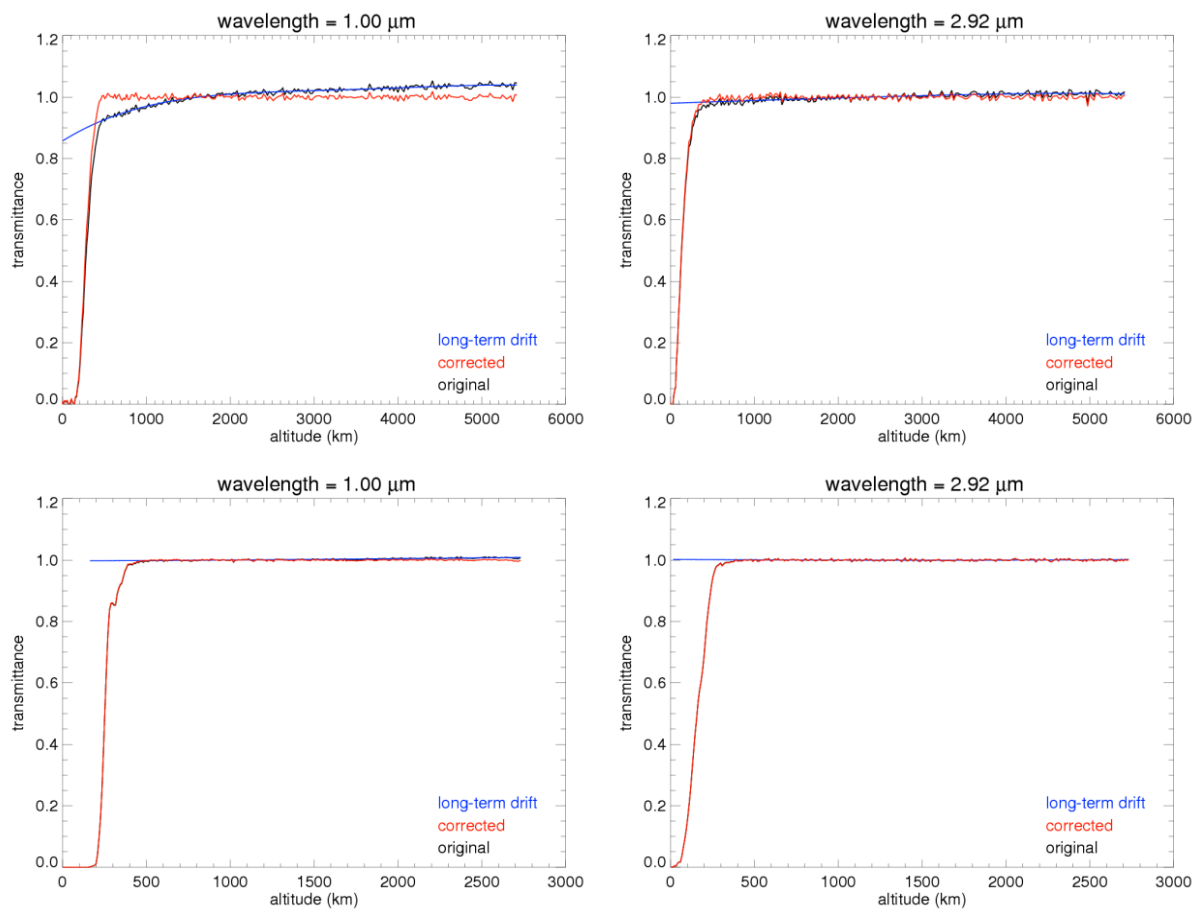

**Figure 8:** Transmittance as a function of altitude at 1.00 µm (left) and 2.91 µm (right) for the T10E (top) and the T78E (bottom) occultations. The behavior of the latter is typical of all the analyzed occultations except T10E. The black and the red curves are respectively the transmittances before and after the polynomial correction, marked in blue. The T78E lightcurves show also that the boresight contamination is greatly reduced.

The uncertainty per spectel has been computed as the standard deviation from the mean transmittance outside the atmosphere, obtained averaging all the spectra above 1000 km. The uncertainty has been calculated separately for each occultation. The regions of the wavelength range where the standard deviation is much higher than the nearest spectels correspond to groups of bad spectels (visible in the left plot of Fig. 7). This happens between 1.62–1.66 μm, between 2.90–3.03 μm and at 3.90 μm, where the filters' edges of the detector lie (Brown et al. 2004). These spectels have been removed from the analysis. This is particularly inconvenient for the 1.62–1.66 μm interval, where a side band of the $CH_4$ band centered at 1.7 μm occurs.

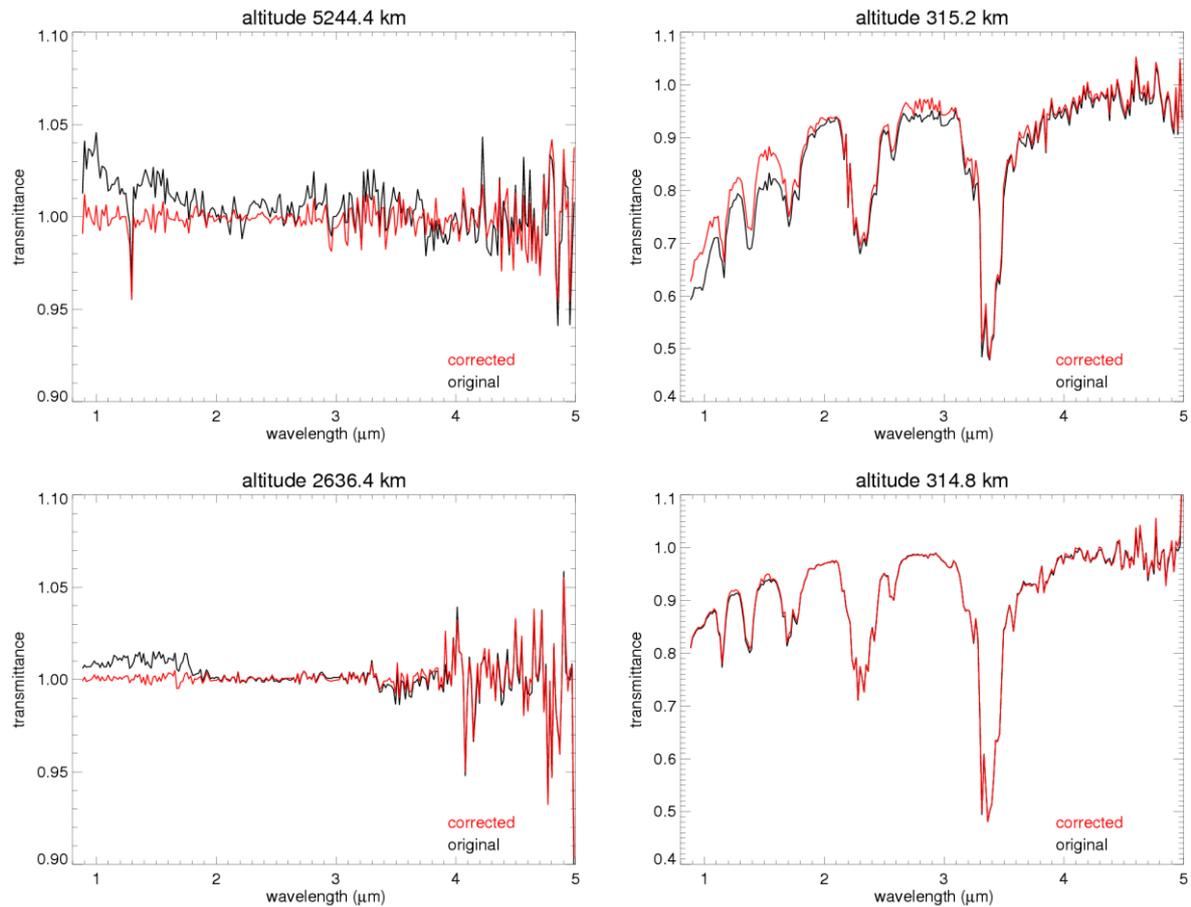

**Figure 9:** Examples of final spectra for the T10E (top) and the T78E (bottom) occultations, one acquired outside Titan's atmosphere (left) and another within (right). The black and the red curves represent respectively the spectra before and after the polynomial correction. The importance of the correction for the T10E occultation is evident, while for the others (T78E included) the application of the polynomial introduces only small variations (but not negligible below ~1.8 μm).

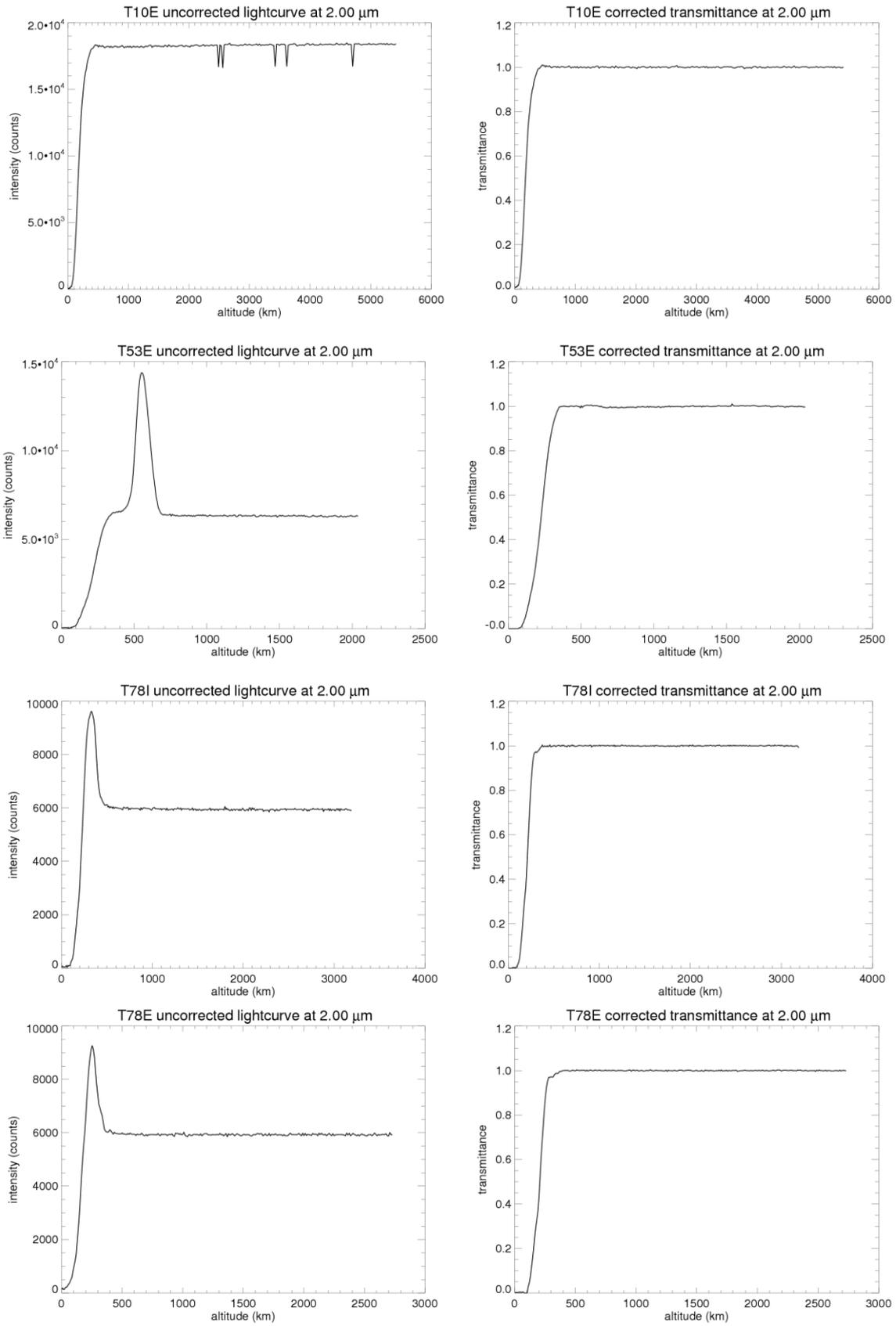

**Figure 10: Lightcurves before the data treatment (left) and final transmission (right) at the 2 μm spectel for all the analyzed occultations.**

Figure 11 shows the transmittances for the four occultations at ten selected altitudes. Five strong bands always appear, at ~1.2, 1.4, 1.7, 2.3 and 3.3 µm, mainly due to methane (with a significant contribution of aerosols for the 3.3 µm band, as discussed in Bellucci et al. 2009). The two strongest bands, at 2.3 and 3.3 µm, are detectable already at ~700 km. At lower altitudes, below ~150 km, the CO band at 4.7 µm appears. An additional signature, previously undetected in Titan's spectra, is present in all occultations around 4.2 µm below ~200 km.

The extinction due to aerosols has the expected strong wavelength dependence, affecting the shorter wavelength first: the spectels around 1 µm reach zero transmittance at ~200 km, to be compared with ~40 km at 5 µm. The aerosol extinction exhibits significant differences between the occultations. T10E is the only one affected by aerosols already at 485 km, but its vertical evolution is very smooth and slow: the T10E spectrum at ~200 km is the least attenuated of the four occultations. This behavior suggests that the haze is very extended vertically but relatively rarefied. The aerosol extinction sets in at lower altitudes in the T53E occultation, but the haze seems to be optically thicker: the spectrum at ~270 km is significantly more affected by aerosols than the spectrum of T10E at the same altitude. The difference between the aerosol's behavior of T10E and T53E, both taken before the equinox, is probably due to latitudinal differences in aerosols' properties more than to seasonal evolution. Indeed the first occultation probes close to the South Pole while the second is at the equator (see Table 1). The large variation in the transmission between ~280 km and 200 km in the two T78 occultations suggests that the haze is mostly concentrated below 280 km and that is very dense, becoming optically thick very quickly. This behavior is consistent with the seasonal evolution of the haze after the equinox, probably due to the change in general circulation (West et al. 2011). This hypothesis is further confirmed by the detection of a detached layer around 310 km in our spectra of the T78 occultations but not in the earlier ones (Fig. 12). This altitude is in good agreement with the observations of the position of the detached layer by ISS measurements around September 2011 (West et al. 2013). Details on the characterization of the haze by VIMS solar occultations will be presented in a forthcoming paper (see Sotin et al. 2013).

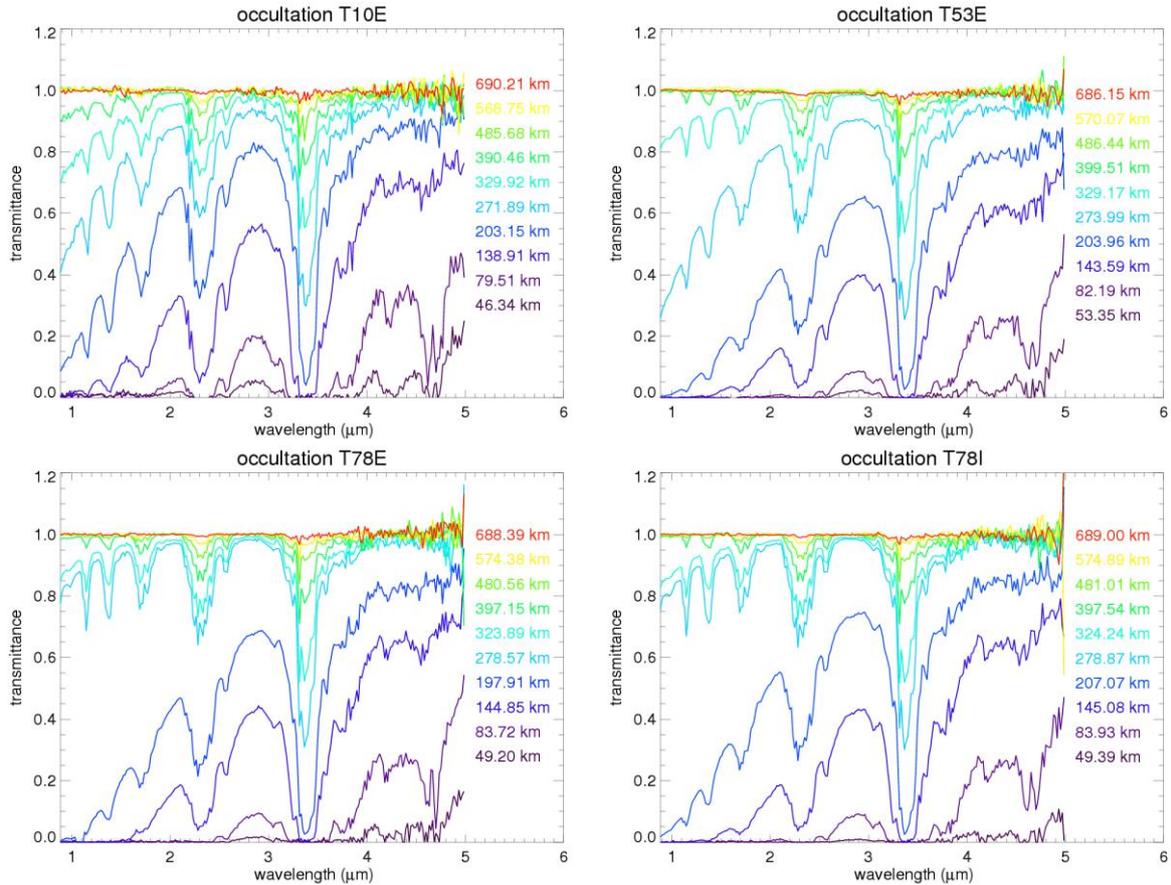

**Figure 11: Examples of 10 transmission spectra per occultation, between 45 and 700 km. The altitudes have been chosen to span approximately the whole atmosphere. The color indicates the altitude of the spectrum according to the legend. Each color corresponds approximately at the same altitude in all the plots.**

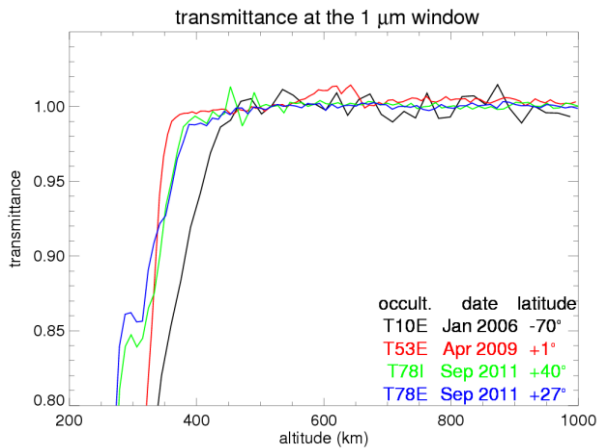

**Figure 12: Transmittance as a function of altitude for the four VIMS occultations at 1 μm, where the absence of gaseous absorptions permits to study the aerosols' behavior. The two T78 occultations exhibit the presence of a detached layer at ~310 km.**

## *3.2 From transmittance to gas vertical profiles*

The transmittance gives information on the total extinction along the optical path at each altitude probed by the occultations, following the Beer-Lambert law. Excluding refraction effects, negligible at the altitudes considered in the present analysis (Bellucci et al. 2009), and atmospheric emissions, not detected by VIMS occultations, the extinction is due to gaseous absorption and aerosol extinction (absorption+scattering). At a given altitude, all the atmospheric layers along the line-of-sight (LOS) contribute to the observed transmittance (slant spectra). In order to extract the vertical profile of atmospheric constituents at the coordinates of the surface nearest point, a spectral and vertical inversion procedure is needed.

The inversion is performed with a Levenberg-Marquardt minimization, based on the optimal estimation methods for nonlinear inverse problems of Rodgers (2000). The minimization procedure is adapted from the one used for SPICAM solar occultations on Mars by Maltagliati et al. (2011, 2013). The main change is its coupling with a different radiative transfer code (Vinatier et al. 2007), which differs from that used by Bellucci et al. (2009) for their analysis of the T10E occultation. Our method consists in extracting the whole profile of the studied atmospheric constituent directly from the slant spectra, then to smooth it with a Tikhonov regularization to minimize the spurious oscillations typical of vertical inversions. We present in this paper the results for the vertical profiles of $CH_4$, using the 1.2–3.0 μm wavelength range, and CO, from its 4.7 μm band (see Sect. 3.2.3 for details). The profiles of the two molecules are inverted separately. This is possible because the CO band is not blended with any $CH_4$ feature.

In order to improve the signal-to-noise ratio at high altitudes where the bands are shallow, we average the spectra by groups of two or three. This procedure decreases the vertical resolution in the highest layers but gives a more stable and reliable retrieval. We applied this averaging technique for altitudes greater than 650 km for methane and 150 km for CO.

For the minimization program we build synthetic VIMS spectra using a line-by-line algorithm. The atmospheric conditions and the spectroscopic coefficients of the molecules included in the model, needed for the creation of the synthetic spectra, are detailed in the next two Sections.

### 3.2.1 Radiative transfer modeling: the atmospheric model

Temperature profiles for the four occultations are provided by CIRS' measurements, interpolated at the altitudes probed by VIMS. The closest CIRS profiles in latitude and season have been employed (Table 2, data presented in Vinatier et al., 2010a and Vinatier et al., submitted). Around the latitude of the T78I occultation (40°N) CIRS shows significant temperature latitudinal trends at the same season (September 2011, after the equinox). Thus, we create a temperature profile by interpolating the two closest CIRS measurements at 29° and 49°N. The CIRS profiles are constrained by measurements between ~$10^{-3}$ and 10 mbar (corresponding to ~500 and ~100 km respectively). We extend the profiles above this limit by taking a constant value corresponding to the last reliable CIRS constraint and at lower levels by employing the HASI profile (Fig. 13).

**Table 2: date and latitude of the CIRS data used for the data treatment. The CIRS flyby used for the equatorial occultation of T53E was obtained two years before the occultation, but at the equator there are no variations between Northern winter (when the CIRS data were acquired) and the equinox season (Vinatier et al., submitted). The T78I profile was extracted by interpolating two CIRS profiles as indicated in the text.**

| Occultation | CIRS data | | |
|---|---|---|---|
| | flyby | date | latitude |
| T10E | T15 | 07/2006 | 54°S |
| T53E | T23 | 01/2007 | 5°N |
| T78I | T76 | 05/2011 | 29°N + 49°N |
| T78E | T76 | 05/2011 | 29°N |

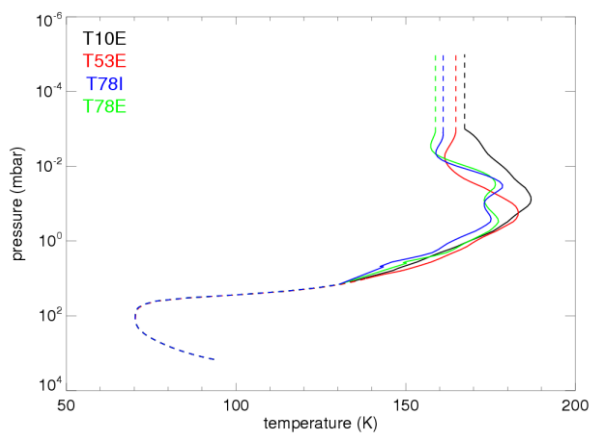

**Figure 13: CIRS temperature profiles used for the four occultations. The dashed parts indicate the extension of the profile outside the altitude range of the CIRS measurements (a constant value for pressures lower than ~$10^{-3}$ mbar and the HASI profile from Fulchignoni et al. (2005) for pressures higher than ~10 mbar).**

We include 9 molecules in the model: $CH_4$, $CH_3D$, $CO$, $CO_2$, $C_2H_2$, $C_2H_4$, $C_2H_6$, $HCN$ and $N_2$. The $CH_3D/CH_4$ ratio is always held constant with a D/H ratio of $1.3 \cdot 10^{-4}$ (Bézard et al. 2007). In the case of $CH_4$ retrieval, the CO value is fixed at the CIRS result of 47 ppm (de Kok et al. 2007). When we invert the CO profile, we use the $CH_4$ profile obtained by our retrieval. $N_2$ is set at a stratospheric mixing ratio of 98.4%. The profiles of $CO_2$, $C_2H_2$, $C_2H_4$, $C_2H_6$ and HCN are taken from CIRS' results (Vinatier et al. 2010a, Vinatier et al. submitted). As for temperature, CIRS retrieves reliable values only for part of the VIMS altitude range (depending on the molecule, CIRS profiles sound from 150 to 300–500 km). We extend the profiles above the CIRS' altitude range with a constant abundance equal to the last reliable

CIRS' value. For the lower atmosphere, we compute the saturation mixing ratio for each gas, using the temperature profile described above, and for each altitude we take the minimum between the mixing ratio at saturation and the last reliable CIRS' value (Fig. 14). We use the same molecules' profiles for the two T78 occultations.

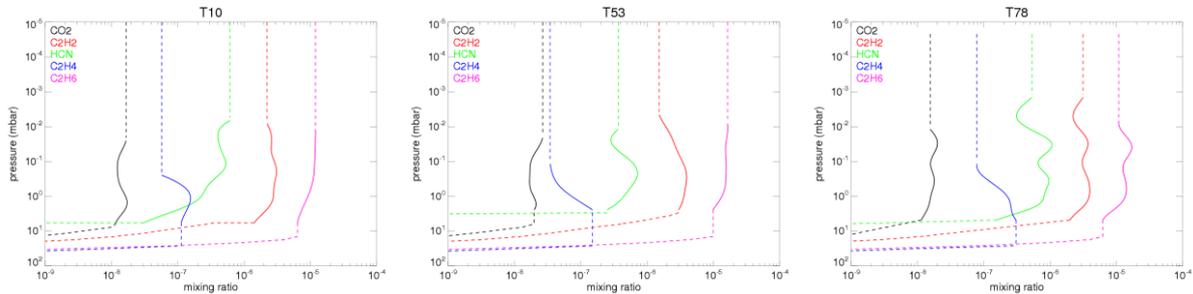

Figure 14: CIRS profiles of the various molecules for the three flybys of the solar occultations. The dashed parts indicate the extrapolation of the profiles outside the CIRS' pressure range (see text for details).

*3.2.2 Radiative transfer modeling: spectroscopic data*

Knowledge of methane's absorption in the near-IR has increased significantly in the last years (Brown et al. 2013). We get our methane coefficients from the merging of three different datasets that represent state-of-the-art knowledge. $CH_4$ lines with $\nu < 5853$ cm$^{-1}$ ($\lambda > 1.708$ μm) were extracted from the VAMDC database (http://vamdc.icb.cnrs.fr/PHP/methane.php) and are based on a global analysis of the methane spectrum as described in Albert et al. (2009) and Boudon et al. (2006). The $CH_4$ lines between 5853 cm$^{-1}$ and 7919 cm$^{-1}$ (1.26 μm < λ < 1.71 μm) are taken or adapted from Campargue et al. (2013 – see discussion below). Finally, for $\nu > 7919$ cm$^{-1}$ (λ < 1.26 μm) the lines are taken from the GEISA 2009 database (Jacquinet-Husson 2011), acknowledging that the triacontad methane band at 1.2 μm is not well modeled yet. For $CH_3D$, we used GEISA 2009 below 3212 cm$^{-1}$ (λ > 3.11 μm), a linelist provided by A. Nikitin (private communication) between 3212 and 5000 cm$^{-1}$ (2 μm < λ < 3.11 μm), based on the works by Nikitin et al. (2002, 2006, 2013), and finally Campargue et al. (2013) above 5853 cm$^{-1}$ (λ < 1.71 μm).

The Campargue et al. (2013) $CH_4$ and $CH_3D$ linelists between 5852 and 7919 cm$^{-1}$ are based on laboratory spectra recorded at 80 and 296 K using Differential Absorption Spectroscopy (DAS) and Cavity Ring Down Spectroscopy (CRDS). While Campargue et al. could determine lower energy levels for lines detected at both temperatures, this is not the case for lines seen at 296 K but not at 80 K. For such lines, Campargue et al. chose lower energy levels such that their intensities are just below the detection limit at 80K. They thus complemented their 80 K linelist with 30282 lines observed only at 296 K. This choice sets an upper limit for methane absorption at temperatures between 80 and 296 K. A second possible approach is to assign to these lines a very high energy level (e.g. 1000 cm$^{-1}$) so that their influence at Titan's temperatures is negligible. This is what was done to generate the HITRAN 2012 linelist (Brown et al. 2013). Campargue et al. (2013) and HITRAN can thus be viewed as yielding respectively upper and lower limits for methane absorption in the 1.26–1.71 µm spectral region for temperatures between 80 and 296 K.

We also made a comparison between VAMDC's and HITRAN 2012's methane lines for the λ > 1.71 µm region and especially the 1.7–2 µm interval, which contains a band of the tetradecad absorption centered at 1.7 µm (Fig. 15). It seems that HITRAN introduces a signature at ~1.8 µm that is not observed in VIMS spectra, and is not present in the VAMDC dataset. It must be noticed that the 1.7–2 µm wavelength range has not been updated in the last HITRAN version (Rothman et al. 2013). Hereafter we use VAMDC+Campargue et al. (2013) as nominal linelist.

The spectroscopic datasets of all the other molecules have been extracted from the GEISA database. We also included the $N_2$-$N_2$ collision-induced absorptions (CIA) using Lafferty et al. (1996) measurements and modeling. It only affects the spectra below ~100 km through the band at ~4.3 µm. The $N_2$-$H_2$ and $H_2$-$H_2$ CIA are negligible in transmission for the altitudes considered in this paper.

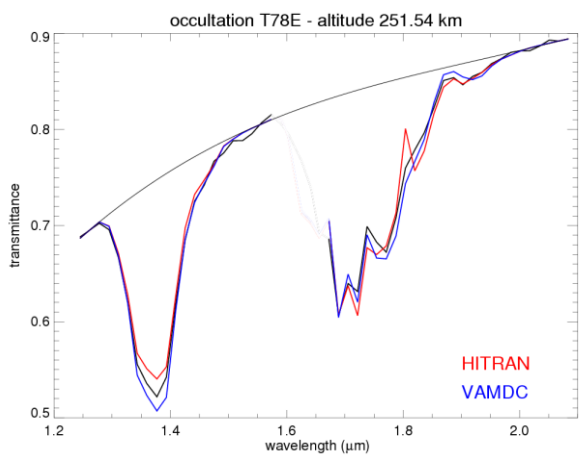

**Figure 15: Example of comparison between synthetic (red: HITRAN 2012; blue: VAMDC + Campargue et al. 2013) and measured (black) spectra, convolved at VIMS' resolution, using two different spectral databases for the methane spectroscopic coefficients. The synthetic spectra have been calculated with a constant 1.4% methane abundance. The 1.82 µm feature observed by HITRAN is not present in the data. The different depth of the 1.4 µm band is a consequence of the difference between the two approaches in treating some lines observed at 296 K by Campargue et al. (2013) spectral lines (see text).**

### *3.2.3 Computation of the slant spectra*

The radiative transfer code divides the atmosphere in concentric spherical layers to compute the slant path spectra. The altitude (and thus the pressure and temperature conditions) of the layers is given in a natural way by the vertical scanning of the occultations. As it is customary for vertical inversions, each layer is supposed to have uniform climatologic conditions and gaseous and aerosol density. The algorithm calculates the line-by-line spectral transmittance of each layer on the line-of-sight, weighted by the optical path of the layer, at every altitude. The last step is the convolution of the hi-res spectrum at the VIMS spectral resolution, taking into account that the FWHM of each spectel varies with wavelength (Sect. 2.1). At higher layers, the same vertical average of the measured spectra (Sect. 3.2) has been applied to the synthetic ones.

Figure 16 shows examples of two synthetic spectra of occultation T78E at two altitudes. Methane (in black) dominates the spectra for most of Titan's atmosphere – even as low as 250 km the contribution of the other molecules is almost negligible. These become significant, at least in parts of the VIMS wavelength range, only in the lowest part of the atmosphere. The spectrum at 126 km shows the strong CO band at 4.7 μm and the peak of $C_2H_2$ (blended with a small HCN contribution) between 3.0–3.1 μm. Features of ethylene and ethane are visible within the strong 3.3 μm $CH_4$ band, but they cannot be disentangled from methane absorption.

A comparison between measured and modeled spectra highlights the presence of a wavelength shift in the three latest occultations (T53E and both T78). Figure 17 shows that the synthetic spectrum can reproduce very well the shape of the 2.3 μm $CH_4$ band for occultation T10E but not for later observations. Agreement can be restored by shifting the whole VIMS spectrum of the T53 and T78 occultations by ~10 nm, around two thirds of the dimension of a spectral channel. We found this value by minimizing the difference between measured and synthetic spectra for different wavelength shifts. The minimum at 10 nm is quite shallow; there is no significant difference if 9 or 11 nm shifts are employed. The presence of a shift is more evident on the shape of the 2.3 μm band but the whole VIMS range is affected. The same shift is thus applied to the whole spectrum. A look at the 2.3 μm band shape in the various occultations seems to suggest that the cause of the shift is a single event happened between T32I (April 2007) and T53E (April 2009). No evolution is observed after T53E, so we apply a 10 nm shift to the data of T53E and both T78. Not all mismatches are eliminated by the shift. In particular, for the 2.3 μm band there are still problems in the shape of the left wing, with the presence of secondary bands in the model that are not appearing in the data.

Our value of 10 nm is quite in good agreement to the wavelength shifts found by Sromovsky et al. (2013) in their analysis of VIMS spectra of Saturn acquired in 2010-11, except for λ > 4.5 µm where the spectra are noisy. Our shift is constant through the whole VIMS range and not wavelength-dependent as in Sromovsky et al. (2013). The hardware configuration of the VIMS instrument is indeed more compatible with a wavelength-independent shift (R. Clark, private communication).

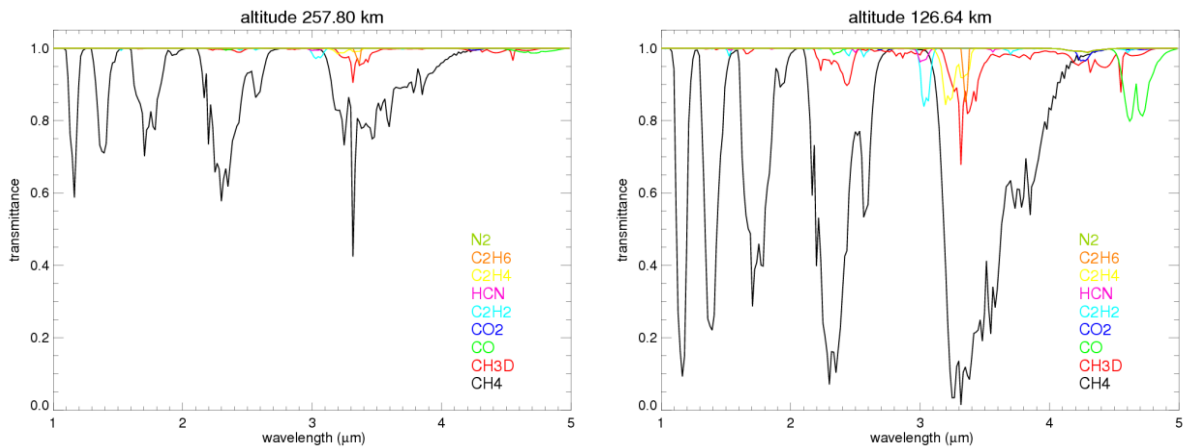

**Figure 16: Model slant spectra at two different altitudes at the climatologic conditions of occultation T10E. Colors indicate the individual contribution of each molecule according to the legend.**

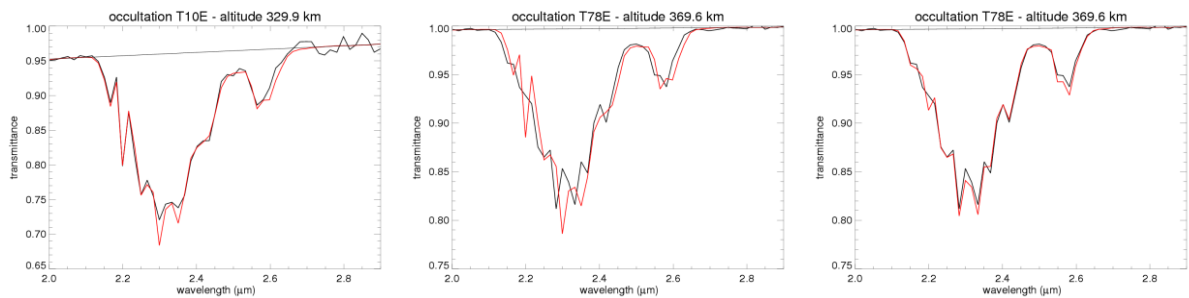

**Figure 17: Synthetic (red) and measured (black) spectra of the 2.3 µm band illustrating the presence of a wavelength shift in VIMS occultations observed after T32I (note: even if the T32I cannot be used for inversion, it allows us to get the overall shape of the 2.3 µm band). Left: T10E occultation. Center: T78E occultation, without shift. Right: the same spectrum of the figure at the center, after applying a 10 nm shift. The synthetic spectrum has been computed with 1.4% of methane. The straight black line on top marks the continuum.**

### *3.2.4 The minimization procedure*

Not all methane bands can be used to extract its vertical mixing ratio profile. The 3.3-µm is strongly contaminated by a series of absorptions attributed to the C–H fundamental stretch of aliphatic chains in aerosol particles (Bellucci et al. 2009), and the spectroscopic coefficients of methane and $CH_3D$ for the 1.2 µm band are not known with sufficient accuracy. Thus, we worked only with the 1.4, 1.7 and 2.3 µm bands. For what concerns CO, only the band at 4.7 µm is sufficiently strong to extract its abundance.

The measured spectra have to be normalized to a continuum. This step also removes the effect of the aerosol extinction from the spectra. The continuum for the methane bands is set at the windows: all the spectels between 1.24–1.28 µm, 1.54–1.58 µm, 2.0–2.1 µm and 2.75–2.9 µm belong to the continuum. In the upper layers of the atmosphere, when the continuum extinction has not started to affect the spectra yet, a linear interpolation is used (Fig. 18a-b). Below we employ a second or a third order polynomial, depending of which fits best the continuum points. Usually, the layers just below the beginning of the aerosol extinction fit equally well with the two options and we choose the simpler second-order continuum (Fig. 18c-d), while a cubic polynomial is necessary for the deeper levels (Fig. 18e-f). The order of the polynomial is sometimes changed ad-hoc after a visual inspection of the spectra in order to take into account some spurious behavior affecting the spectra in the continuum intervals, as for example residuals from the main boresight contamination. Each occultation has its own set of continuum polynomials. The same polynomials are applied to the synthetic spectra. Because the continuum windows are not completely transparent at low altitudes but contain a small contribution from methane (see the right plot of Fig. 16 as an example), this "pseudo-continuum" is recomputed at each iteration of the minimization procedure, when the $CH_4$ abundance is updated.

In order to take into account the uncertainties associated with the continuum choice, we use different wavelength ranges for the inversion, each with its own continuum: 1.24–2.01 µm, 1.54–2.90 µm, and 1.24–2.90 µm. For each wavelength range we extract several methane profiles by fitting the whole wavelength range and each methane band separately (for example, from the 1.24–2.01 µm interval we retrieve one profile from the inversion of the whole range and two more by fitting independently the two bands included in it, at 1.4 and 1.7 µm). This means that for each occultation we have 10 different methane profiles. The results and the comparison between the profiles extracted by the various wavelength combinations are presented in Section 4.1.

For the 4.7 µm CO band the continuum choice is more straightforward. The 4–5 µm range of VIMS is noisy and we applied a simple linear continuum using the 4.13–4.17 µm and 4.90-4.96 µm intervals as continuum points. The left continuum is affected by non-negligible absorptions of $CH_4$ and $CH_3D$, which are directly included in the continuum.

The Levenberg-Marquardt method needs to set an a priori of the retrieved profile. We employed a constant abundance of 1.48% for methane (the GCMS result, Niemann et al. 2010) and 47 ppm for CO (the CIRS value, De Kok et al. 2007). We verified that the final profiles are insensitive to the a priori. We set six iterations per retrieved mixing ratio profile.

Usually the fit is already stable after the first two iterations, but we kept six iterations to take into account the most difficult cases. The uncertainty in the density profiles due to the inversion is given by the covariance matrix of the solution errors. The other main contribution to the final error is the propagation of the uncertainty on temperature profiles, which is less than 1 K. To be conservative, we run some tests by changing the temperature profile by ±1 K. This causes a corresponding absolute uncertainty on the order of ~0.07% on the methane mixing ratio, regardless of the occultation and the fitted band. The final error on methane is given by the quadratic sum of these two contributions. The CO retrieval is instead almost insensitive to temperature thus its error is only the one given by the inversion.

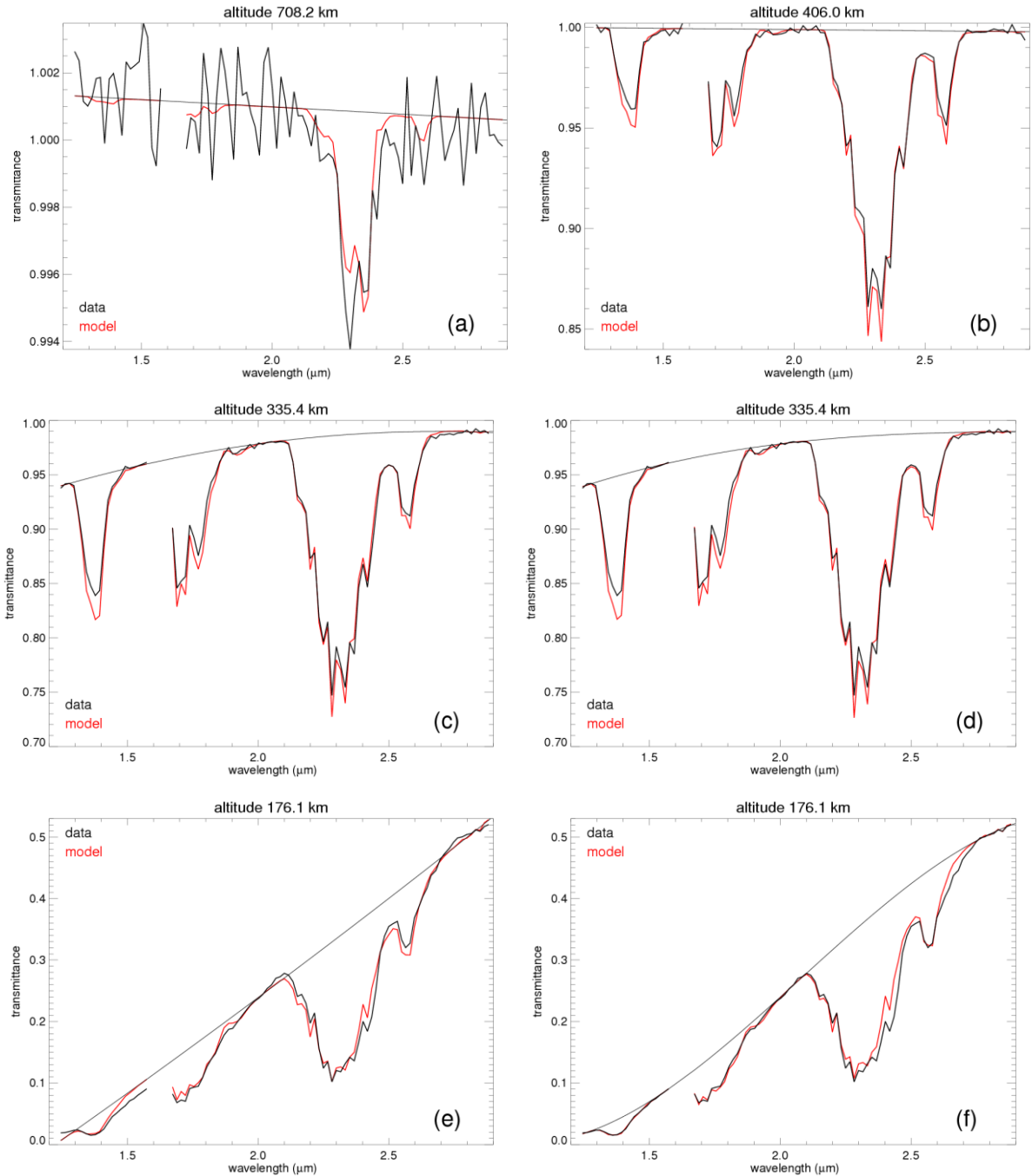

Figure 18: Examples of continuum choices comparing data (thick black curve) and synthetic spectra computed with 1.4% of methane (red curve) normalized at the same continuum points, for the 1.24-2.9 μm wavelength range. The continuum is the thin black curve. Top row (a-b): linear continuum before the aerosol extinction starts to affect the spectra. (a): occultation T78I. This plot also shows that VIMS occultations can detect the 2.3 μm methane band even when the band depth is just ~0.005, i.e. at altitudes as high as 700 km. (b): occultation T53E. The rest of the spectra are from T53E. Middle row (c-d): just below the beginning of the aerosol's extinction, we cannot discriminate if it is better to fit the continuum with a second-order (c) or third-order (d) polynomial. In these cases the simple parabolic continuum is chosen. Bottom row (e-f): in the lower part of the atmosphere the second-order continuum (e) cannot reproduce the shape of the wavelength dependence of the extinction and the third-order continuum (f) is favored.

## 4 Results

### 4.1 Methane

Figure 19 shows all the 10 methane mixing ratio profiles retrieved for the T78I occultation. Each panel pertains to a different $CH_4$ band, and the various profiles within each plot represent different continuums choices. The multi-band profiles are also shown. All the profiles from the same band are in good agreement with each other for most of the altitude range, proving that the result is quite robust with respect to the continuum choice. The profiles diverge the most at the highest and lowest layers of the atmosphere. This happens for two different reasons. In the upper atmosphere the band depth is comparable with the spectral noise, thus fluctuations in the continuum windows have a bigger impact in the determination of methane's abundance than at lower altitudes. This is reflected by the big error bars. In the lower atmosphere, instead, the increasing spread between the profiles is due to the strong aerosol extinction in the 1.2–1.7 µm range, which tends to saturate the bands and increases the sensitivity to the shape of the continuum employed for the fit. The multi-band fits do not add any significant information.

The similarity between the profiles derived from a given band is present on all the four occultations, so we can use an average profile for each band. The average is computed with the error bars as weights. The dispersion of the profiles at each altitude is used to define the uncertainty of the averaged profile. If the dispersion is less than the largest error of the individual profiles, we take the latter instead.

Figure 20 shows the average profiles for the four occultations. Retrievals for the T53E are spoiled above 470 km because of the residuals of the main boresight contamination. The general behavior is consistent throughout the whole dataset. The mixing ratio profiles from the 1.4 and 1.7 µm bands are always in agreement with each other, with an approximately constant abundance of ~1.2% up to ~450 km, lower than the GCMS reference value of 1.48%. There is the tendency in both profiles to join the GCMS abundance in the upper part of the atmosphere. The 1.4 µm band becomes too shallow above 500 km so its retrieval, with big error bars, is just indicative. The biggest difference between the two profiles is in the lowest atmospheric layers, mainly due to the effect of the aerosol extinction described above. The profiles however usually overlap within the error bars even in the lowest atmosphere. The profiles from the 2.3 µm band have instead a completely different behavior. They are in good agreement with the GCMS value in the upper atmosphere. Then, they exhibit a minimum around 400 km before increasing to values much higher than GCMS below ~300 km. The 2.3-µm profile of T10E was the one presented by Bellucci et al. (2009). Our results are in good agreement with theirs, especially for what concerns the puzzling increase at low altitudes.

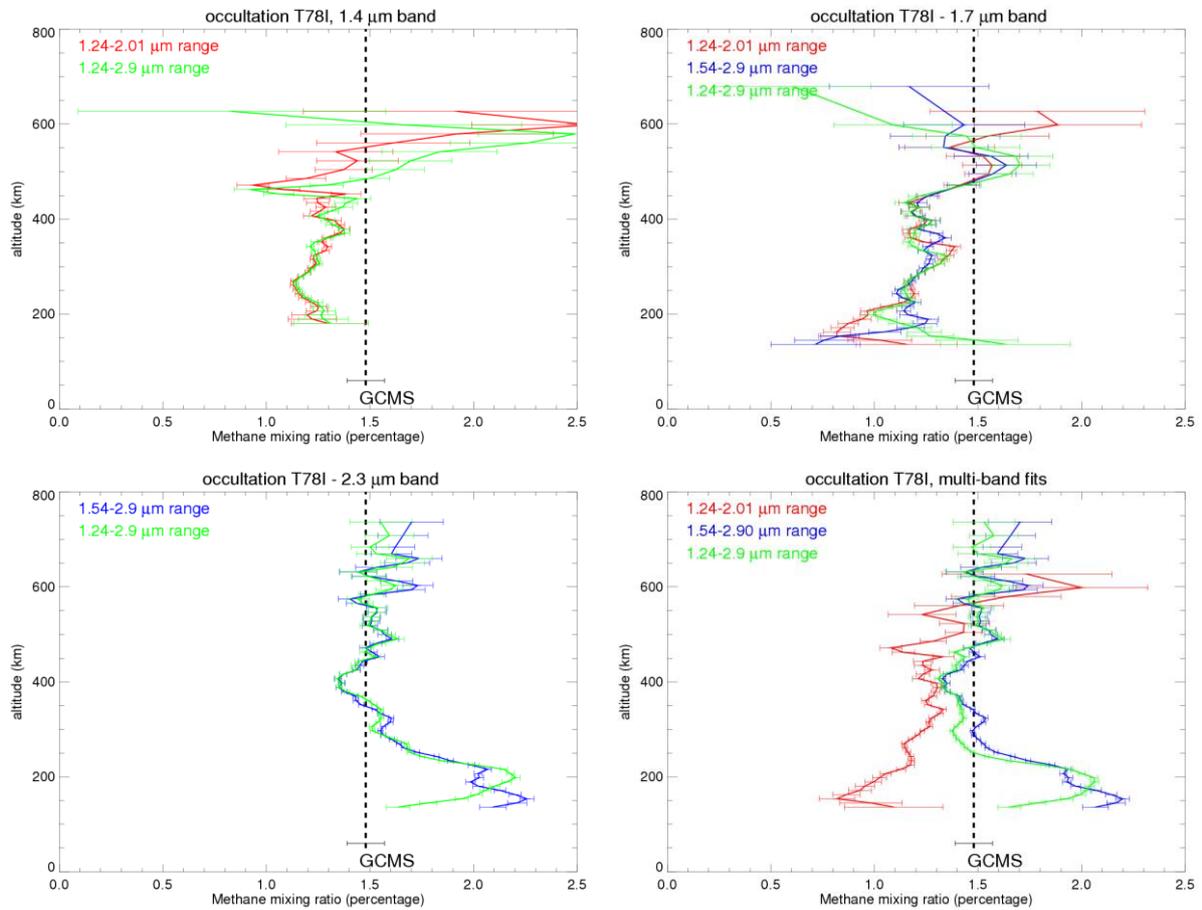

Figure 19: Methane abundance profiles retrieved from occultation T78I. Each plot regroups the profiles extracted from the same band (top left: 1.4 µm; top right: 1.7 µm; bottom left: 2.3 µm) except the bottom right which shows the profiles obtained by fitting simultaneously more than one band (1.4+1.7 µm bands for the 1.24–2.01 µm wavelength range; 1.7+2.3 µm bands for the 1.54–2.90 µm range; and 1.4+1.7+2.3 µm bands for the 1.24–2.9 µm range). The color identifies the wavelength range (and thus the continuum) used for that profile, according to the legend present in each figure. The error bars show the uncertainty from the Levenberg-Marquardt inversion. The vertical dashed line marks the GCMS reference abundance of 1.48%, with its error bar of 0.09% (Niemann et al. 2010).

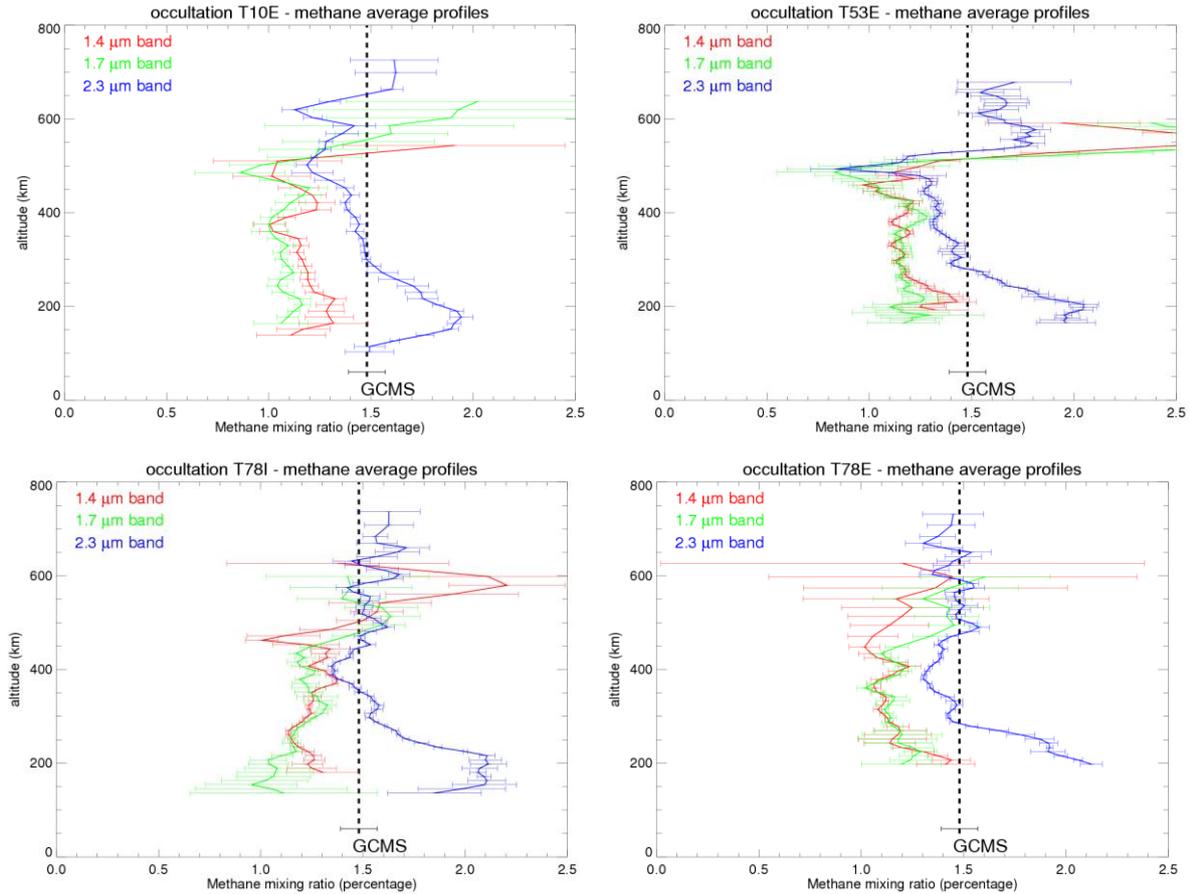

**Figure 20:** Average methane abundance profiles from each band for the four occultations. The color identifies the band, according to the legend. The reference GCMS is marked by the dashed vertical line.

We cannot use the 2.3-µm profile as reference for methane's abundance, because its behavior with altitude is unphysical at Titan's conditions and it yields unrealistic values. In fact, methane photolysis occurs on Titan only above 700 km, and methane enrichment from surface sources affects only the first 30 km of the atmosphere, so in the altitude range observed by VIMS methane should be well-mixed. Our interpretation of the behavior of the 2.3-µm profile will be discussed in detail in Sect. 4.3. The profiles from the 1.4 and 1.7 µm bands are in very good internal agreement, thus their result seems to be more reliable. However, as they are significantly lower than the GCMS value, we investigate the possibility of a bias towards low $CH_4$ values in our procedure.

A possible bias could come from our choice of the methane coefficients for the 1.26 µm < λ < 1.71 µm interval. In fact, as described in Sect. 3.2.2, the Campargue et al. (2013) treatment of those $CH_4$ lines seen at 296 K but not at 80 K by the DAS/CRDS instrument gives an upper limit for methane absorption in that region. Conversely, the HITRAN 2012 approach of the same wavelength region gives the lower absorption limit (see also Fig. 15). The "real" methane absorption, thus the "correct" methane profile, is somewhere in between these two extremes. We tested the impact of the difference of the two linelists on the methane profile obtained from the inversion of the 1.4 µm band (Fig. 21). The discrepancy between the two profiles is remarkable especially below 350 km. As expected, the HITRAN 2012 methane coefficients yield a more abundant profile. However, its increase in the lower atmosphere and its very high abundances, especially below 200 km, are not physical at Titan's conditions. In addition, the profiles obtained with the Campargue et al. (2013) data are in very good agreement with the inversions from the 1.7 µm band, whose spectral coefficients come from another independent source. We thus maintain the Campargue et al. (2013) linelist, which we believe does not introduce significant biases in the determination of the methane mixing ratio profile, within the uncertainties. It must be noted that Bézard (2014) arrives at the opposite conclusion and proposes a correction to the energy level of Campargue et al. for those methane lines observed at 296 K only. However, Bézard (2014) analyzes just the 1.4 µm methane band, observed by the Upward Looking Infrared Spectrometer of the Descent Imager Spectral Radiometer (DISR/ULIS), and thus it cannot make any comparison with the 1.7 µm band. Spectroscopic laboratory measurements of methane at Titan's temperatures could give the definitive answer, while for now this remains an open question.

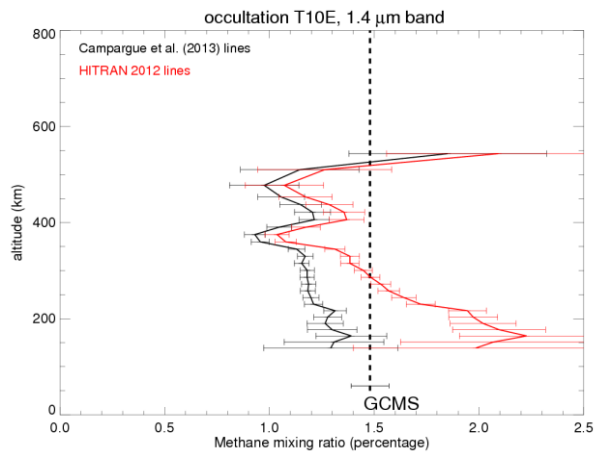

**Figure 21: Comparison of methane mixing ratio profile extracted from the 1.4 µm band of occultation T10E using the Campargue et al. (2013) linelist (black) and the HITRAN 2012 linelist (red). For this test we used the 1.24-2.01 µm wavelength range as continuum reference. The other occultations exhibit a similar behavior.**

An alternative interpretation is that we do not include in our model some physical phenomena that are present in Titan's atmosphere, resulting in an overestimation of our synthetic methane bands. The best candidate is forward-scattering. Aerosols on Titan are strongly forward-scattering (Tomasko et al. 2008b), but it is not straightforward to understand how they affect the methane bands depth in transmission. This effect has been investigated by De Kok & Stam (2012). They showed that, in occultation geometry, ignoring forward-scattering in the spectral model of the 2.3 µm band underestimates the volume mixing ratio of methane by ~8%. It is interesting to notice that their retrieved profile bends above ~350 km and approaches the "real" CH$_4$ abundance in the upper atmosphere (Fig. 22). This behavior is similar to what we observe for the 1.4 and 1.7 µm bands, and perhaps for the 2.3-µm as well, if we consider the minimum that appears in its profiles at ~400 km as a real feature and the increase in the lower atmosphere as an effect due to another phenomenon. The same computation performed for the 1.4 µm band shows that the underestimation is even stronger at lower wavelengths, increasing from 8% to 12% (Fig. 22). Changes in the aerosol vertical distribution as those observed by our occultations (see Fig. 11) do not have a significant impact on this correction. However, we observe a much stronger decrease, of ~20% with respect to the GCMS value. Moreover, the correction for the 1.4-µm shows an almost constant profile with no mixing ratio increase at high altitudes, present instead in our profiles. This behavior suggests that, while neglecting forward-scattering can affect our estimation of methane abundance, it cannot fully explain the observed discrepancy with GCMS. Recent studies have put into question our knowledge of methane stratospheric abundance and currently there is no general agreement. An analysis of CIRS data found values as low as 1% at some latitudes (Lellouch et al. 2014), while the DISR/ULIS measurement analyzed by Bézard (2014) confirms the GCMS results. Our results fall between these two measurements.

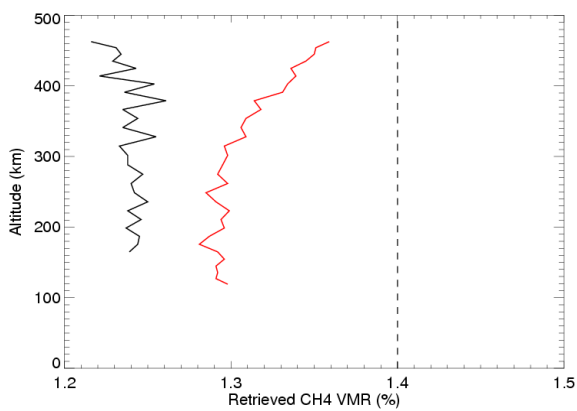

**Figure 22: Methane volume mixing ratio profile retrieved from the 1.4 µm band (black solid line) and from the 2.3 µm band (red – same profile presented in Fig. 6 of De Kok & Stam, 2012) by ignoring forward-scattering by the aerosol with respect to the value used to build the simulated data (dashed line). The impact of neglecting the forward-scattering is stronger for the 1.4-µm, since the aerosols are more forward-scattering at shorter wavelengths and the optical thickness is larger as well.**

To conclude, we take as the inferred CH$_4$ value for each occultation the average of the profiles derived from the 1.4 µm and the 1.7 µm band inversions, weighted by the uncertainty at each altitude. We then increase the result by 10% to account approximately for the neglect of forward-scattering in the model. Results are summarized in Table 3.

Table 3: Reference $CH_4$ and CO values from the four VIMS solar occultations. The absolute uncertainties of methane's abundance after the forward-scattering correction are the same of the uncorrected values.

| Occultation | $CH_4$ abundance (%) | $CH_4$ abundance (%) after forward-scattering correction | CO abundance (ppm) |
|---|---|---|---|
| T10E | 1.14±0.08 | 1.25 | 30.6±5.1 |
| T53E | 1.15±0.07 | 1.27 | 57.0±8.6 |
| T78I | 1.21±0.07 | 1.33 | 61.1±8.4 |
| T78E | 1.12±0.08 | 1.23 | 36.8±5.8 |

*4.2 Carbon monoxide*

CO is difficult to retrieve with precision because its band belongs to the noisiest part of VIMS' spectra. In particular, its right-side continuum, between 4.90–4.96 µm, exhibits strong random fluctuations. This explains the large error bars in the retrieval of its profile. Nevertheless, the CO band is clearly identified between 60 and 160 km. Figure 23 shows examples of spectra of the 4–5 µm region at different altitudes for the T53E occultation. The other occultations exhibit similar trends. The band that appears in the synthetic spectra between 4.2–4.35 µm below 100 km is due to the $N_2$-$N_2$ CIA and becomes dominant in that part of the spectrum below 75 km. $CO_2$ is also present in the model at the same wavelengths (Fig. 15), but its contribution is minor with respect to the $N_2$-$N_2$ CIA.

In addition to the CO band, the most striking feature in this spectral range is the discrepancy between the data and the modeled spectra between 4.15–4.5 µm. The VIMS spectra suggest the presence of additional absorptions in that wavelength range, roughly consisting of a separate band centered at 4.2 µm plus a broader absorption between 4.3–4.5 µm. These features are consistent between all the occultations and in their vertical evolution. They also are well outside the transmittance's uncertainties of VIMS data, making their detection secure. Their behavior and possible identification are discussed in the next Section.

Figure 24 shows the retrieved CO profiles from all the occultations. The error bars of the profiles are most probably underestimated because they do not fully take into account the uncertainties on the continuum slope. The profiles are vertically quite constant and in good agreement with the CIRS value of 47 ppm. The overall average CO value between all the occultations is indeed 46±16 ppm; this relative error of ~30% is probably a better indication of the uncertainty of our CO retrievals. Our results are also in agreement, within the errors, with the stratospheric abundance of 32±15 ppm found by VIMS nightside limb measurements (Baines et al. 2006). CO is supposed to be very stable and well-distributed in Titan's atmosphere, because it has a lifetime of ~500 Myr (Lellouch et al. 2003), much higher than the mixing timescale, and it is a non-condensable gas.

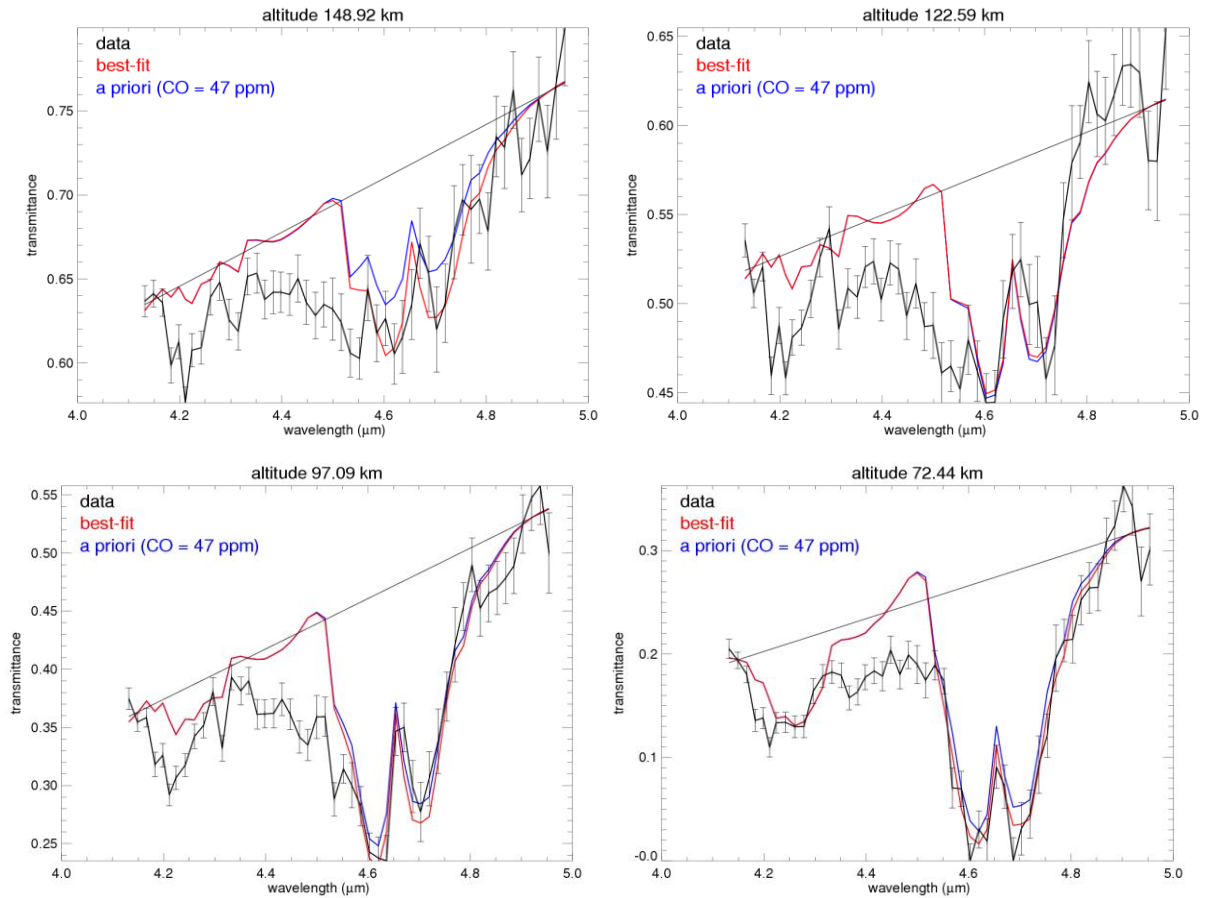

**Figure 23: Examples of spectral fits of the CO band at 4.7 μm from the T53E occultation. Black: data, with errors. Blue: synthetic spectrum computed at the beginning of the inversion procedure, with the a priori value of 47 ppm given by CIRS (de Kok et al. 2007). Red: best-fit synthetic spectrum. The thin black straight line marks the linear continuum. Below 100 km the $N_2$-$N_2$ CIA absorption is visible at ~4.25 μm.**

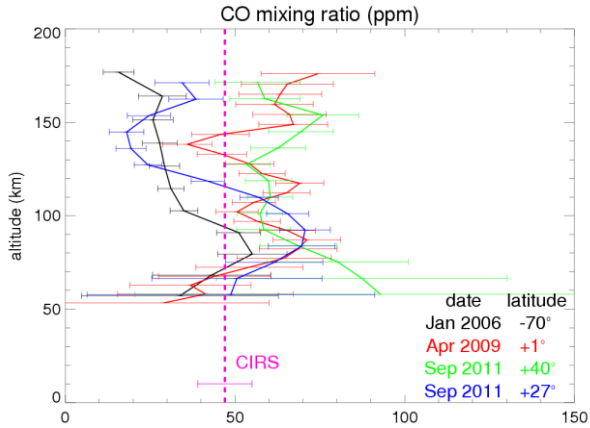

**Figure 24**: CO volume mixing ratio vertical profiles retrieved from the four occultations (labeled by color). The pink dashed vertical line marks CIRS' stratospheric CO abundance (de Kok et al. 2007).

*4.3 Other absorption bands*

VIMS transmission spectra exhibit additional features that cannot be explained with the molecular absorptions included in our model. The strongest of them is the broad absorption centered at 3.4 μm that was identified by Bellucci et al. (2009) as the signature of the C–H stretching mode of aliphatic hydrocarbons. The origin of these modes has been attributed by Bellucci et al. (2009) to the haze, but other gaseous contributions could be present. Rannou et al. (2010) showed that the absorption is too deep to be generated exclusively by the haze, and Kim et al. (2011) identify some secondary peaks as signatures of hydrocarbon ices. Bellucci et al. (2009) suggest that additional absorptions are also present within the 2.3 μm methane band. This is the most straightforward and reasonable explanation for the increase of the retrieved methane from the 2.3 μm band at low altitudes. Contrary to the 3.4 μm band, however, here the additional absorption is minor with respect to the methane band, and this complicates the identification. Bellucci et al. (2009) do not give a definite answer but suggest that the same haze creating the 3.4 μm could be at the origin of the absorption at 2.3 μm through its weaker combination bands. Sim et al. (2013) show that the residual absorptions at 2.2–2.4 μm and 3.3–3.5 μm can be qualitatively fitted with the same set of hydrocarbon ices. Also Kim & Courtin (2013) interpret the additional absorption features they find in the 1-5 μm wavelength range from VIMS solar occultations as generated by alkane ices.

Our dataset allows us for the first time to study the shape and vertical behavior of these additional absorptions in more than one occultation and in different seasons. We perform a direct calculation for each occultation by imposing a methane abundance equal to the average value indicated in Table 3 and we extract the residual spectrum as a ratio between the measured and the synthetic spectra, normalized to the same continuum (Fig. 25). Figure 26 presents the residual spectra for all four occultations for the 1.54–4.0 µm wavelength range, which contains the three strongest methane band of the VIMS range.

As Fig. 26 shows, the shape and vertical evolution of the residual absorption is very consistent between the four occultations, further proof that we are detecting real features, the strongest of them being by far the one at 3.4 µm. This broad band, which saturates below ~160 km, has a very complex structure. For the T10E occultation, the main absorption peak is at 3.38 µm and there are indications of the presence of secondary peaks at 3.41 and 3.45 µm (for later occultations, due to the wavelength shift of VIMS' spectra discussed in Sect. 3.2.3, the peaks are shifted by one spectel towards lower wavelengths). As Bellucci et al. (2009) remark, this is in very good agreement with the peaks of the asymmetric C–H stretch of –$CH_3$ and –$CH_2$ groups observed by Sandford et al. (1991) and D'Hendecourt & Allamandola (1986). Towards longer wavelengths, two secondary absorptions within the extended wing are seen at 3.6 and 3.75 µm. It is more difficult to interpret these absorptions as the product of C–H stretching transitions. The only C–H stretching bands in this wavelength range are derived from aldehyde, whose presence on Titan's atmosphere is improbable.

The left wing of the 3.4 µm feature is much steeper, but it exhibits a distinct isolated peak at 3.28 µm. This is the wavelength where a strong and narrow absorption due to aromatic molecules as the polycyclic aromatic hydrocarbons (PAHs) is detected in many different astronomical objects as well as in the upper atmosphere of Titan (Lopez-Puertas et al. 2013).

The absorption around 2.3 µm is the strongest signature outside the 3.2–3.8 µm range (Fig. 26). It starts to be detected at ~400 km and can account for almost 10–15% of the absorption at 2.4 µm around 200 km. It exhibits a distinct two-lobed shape, with one lobe between 2.2–2.35 µm and the other between 2.35–2.5 µm, with a clear minimum in between. A weaker feature is present between 1.65–1.75 µm, at the center of the 1.7 µm $CH_4$ band. Also this band seems to have a double-peaked shape with a minimum in between at the 1.70 µm spectel. Another absorption of comparable intensity is detected around 3.1 µm. Finally, there are two minor but visible signatures, centered at ~1.9 and ~2.65 µm.

The residuals exhibit also some small wavelength intervals where the synthetic spectrum is actually deeper than the measured one. The zigzag shapes around 2.2 µm are due to a bad reproduction of the left wing of the 2.3 µm band for those occultations where the wavelength shift is applied. In fact we do not see such structures in the T10E occultation. The other region where the synthetic spectrum is more absorbing than the measured spectrum is around 3.2 µm. It is a range with a steep slope of the methane band blended with a significant $C_2H_4$ absorption, and even small errors in the molecular abundances can have a significant impact in the spectral shape.

Many of these absorptions have been detected for the first time. Through the analysis of their shape and evolution with respect to altitude we have found a good candidate for their identification: gaseous ethane. Ethane is the second most abundant hydrocarbon on Titan (after methane). However, its near-infrared spectrum is complex and very difficult to model, so a database of reliable $C_2H_6$ spectroscopic parameters in the VIMS' range currently does not exist. The two main collections of atmospheric coefficients, HITRAN and GEISA, contain only a few lines between 3.3–3.4 µm, which can be seen in the right plot of Fig. 16. However, even this very partial coverage is not complete, because the theoretical computations include only the nine strongest multiplets of $C_2H_6$ in that region and neglect the complex underlying envelope of weaker lines. Nevertheless, laboratory measurements of the near-IR spectrum of ethane have been obtained by the Pacific Northwest National Lab (PNNL – Sharpe et al. 2004). Figure 27 shows a comparison between the VIMS residuals and the PNNL measurements in the 1.54–4.0 µm wavelength range. Most of the features exhibit a very good agreement. Not only the bilobate shape of the 2.2–2.5 µm band is excellently reproduced, as well as the relative intensity of the two lobes, but almost all the secondary bands of VIMS residuals (at 1.7, 2.7, 3.1, as well as the above-mentioned 3.6 and 3.75 µm bands) have a corresponding match in the laboratory spectra. The intensity proportion between them is also respected, with the possible exception of the 1.7 µm residual which is stronger in the VIMS data with respect to the other signatures. The only residual band that does not seem to be due to ethane is the peak at 1.9 µm.

The other region of the VIMS spectrum where we find additional absorptions is the 4.1–4.5 µm range: a well-defined one at 4.2 µm and a broader one between 4.3–4.5 µm (Sect. 4.2). Here we could not create a good residual spectrum as a function of altitude like those of Fig. 26 because of the high noise level beyond 4 µm (Fig. 28). However, we can still extract some qualitative information. Both these absorptions start to be detected in the lower layers of the atmosphere, the former below ~230 km and the latter below ~180 km. We could not find a good candidate for the 4.2 µm band. It is difficult to attribute it to a stretching mode linked to the haze, because there are no absorptions due to stretching variations around this wavelength. Regarding the gases, the most probable candidates are included in our atmospheric model. The $CO_2$ band at ~4.25 µm is not deep enough and it is also slightly displaced towards longer wavelengths with respect to our absorption. Again, ethane is a possibility. As Fig. 28 shows, PNNL measures an absorption band between ~4.17–4.22 µm with the peak at the right place. However, it seems too weak (maximum cross section $2 \cdot 10^{-21}$ cm$^2$/molecule) to produce such a strong absorption. Moreover, the other (weaker) branch of the ethane band that peaks at 4.27 µm does not seem to be clearly detected. No other gases in the PNNL database exhibit clear features at 4.2 µm. We also notice that the depth of this residual band does not seem to grow below ~90 km, so that the $N_2$-$N_2$ CIA, which appears below 100 km centered at ~4.25 µm, fills slowly up the residual band (this effect can be seen clearly in the 72 km spectrum of Fig. 23). This behavior could be consistent with an absorption band generated by a gaseous component whose abundance decreases sharply below ~90 km, perhaps due to condensation. The 4.3–4.5 µm is another region with no strong distinct signatures. Ethane has an envelope of many weak lines that is probably not sufficient to create the absorption we observe. This wavelength range contains however several absorptions due to stretching variations especially between 4.4–4.55 µm, particularly a sharp feature of nitriles from the –C≡N stretch and several weak bands from the –C≡C– stretch of alkynes. All these effects combined together could create a significant signature.

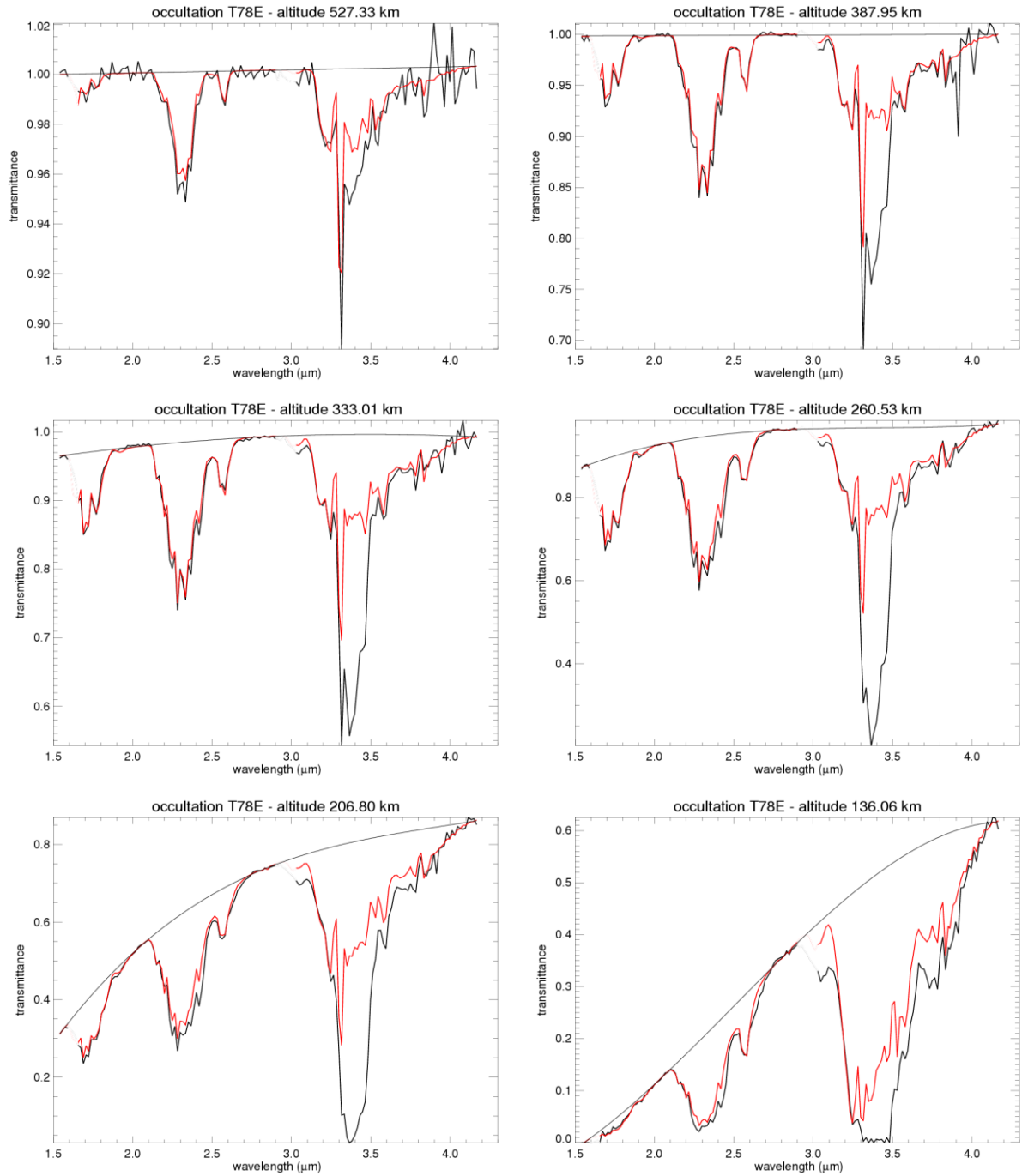

Figure 25: Comparison of synthetic spectra (red) computed at different altitudes with spectra observed during the T78E occultation. The synthetic spectra are normalized at the same continuum of the VIMS data (black). The continuum is indicated by the thin black line on top. The wavelength range is the same as in Fig. 26.

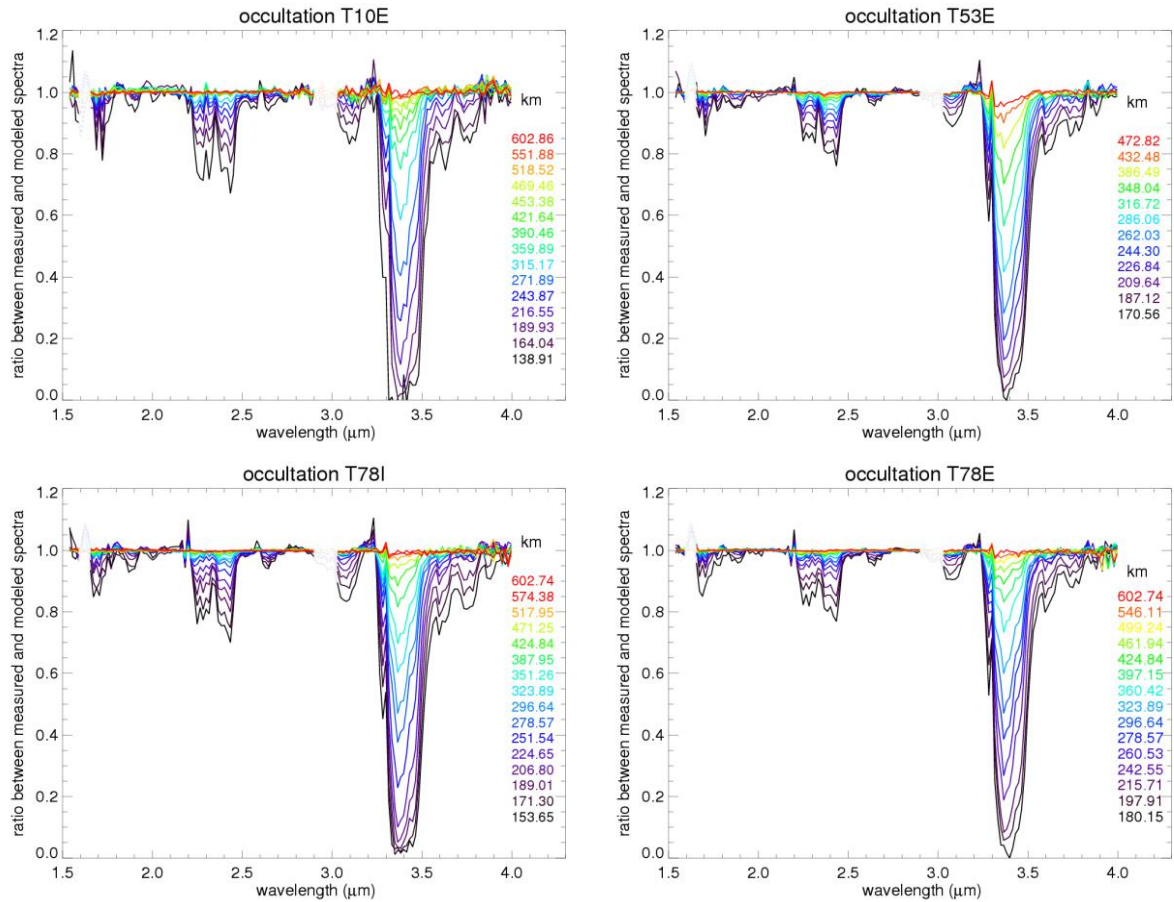

Figure 26: Altitude evolution (in color according the legend) of the residual spectra for the four occultations, between 1.54–4.0 µm. The synthetic spectrum has been computed with the mean methane abundance corresponding to each occultation (Table 3). Note that the T53E plot is less extended in altitude because the residuals of the parasitic light from the main boresight compromises the residuals above ~480 km. Some fluctuations at λ > 3.8 µm are due to an increase of the noise in that part of the spectrum, which affects the continuum determination.

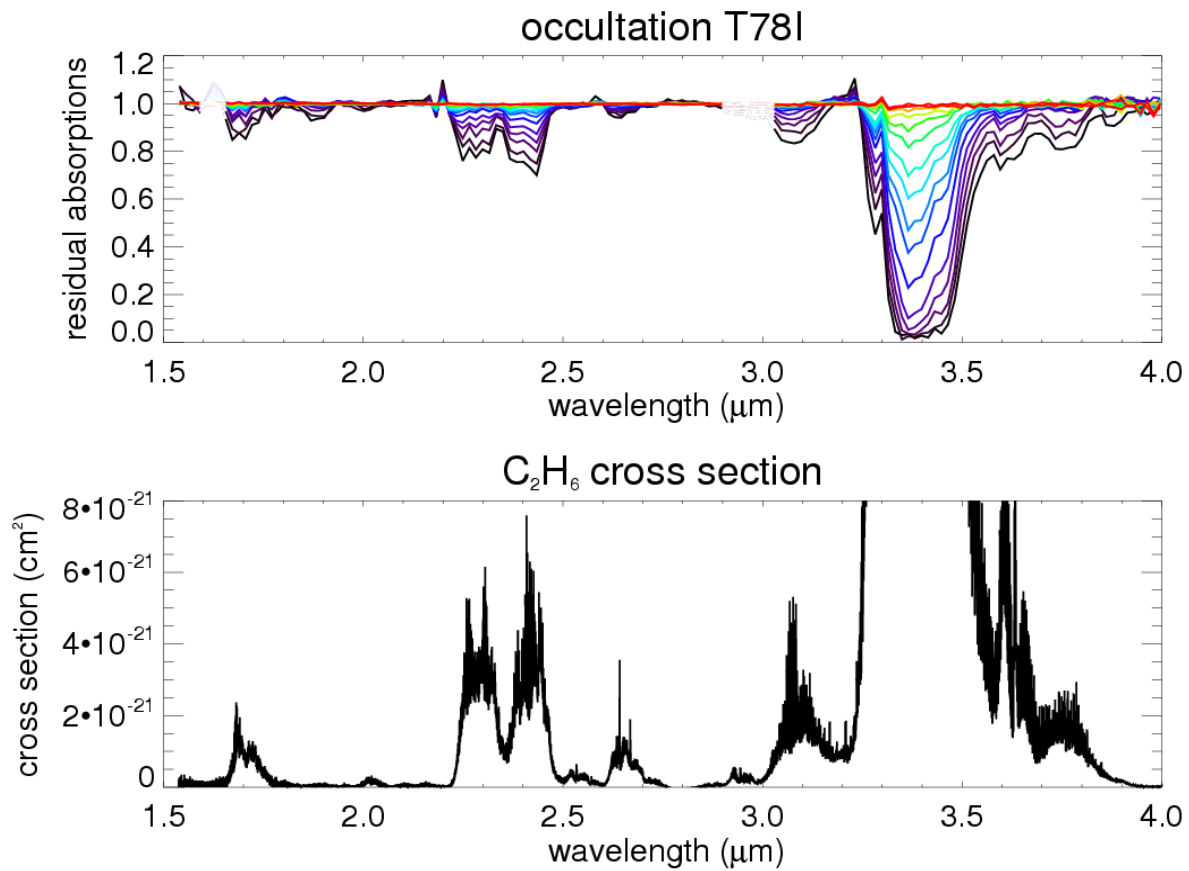

**Figure 27:** Comparison between the PNNL ethane spectrum obtained at 5°C (bottom plot) expressed in cross section (cm$^2$/molecule) and the additional absorptions found by VIMS for the T78I occultation (top, same plot as Fig. 25) in the 1.54–4.0 μm wavelength range. Its vertical scale is set to highlight the weaker bands outside the 3.25–3.55 μm range. The ethane cross section between 3.25–3.55 μm can be as high as 2·10$^{-18}$ cm$^2$/molecule.

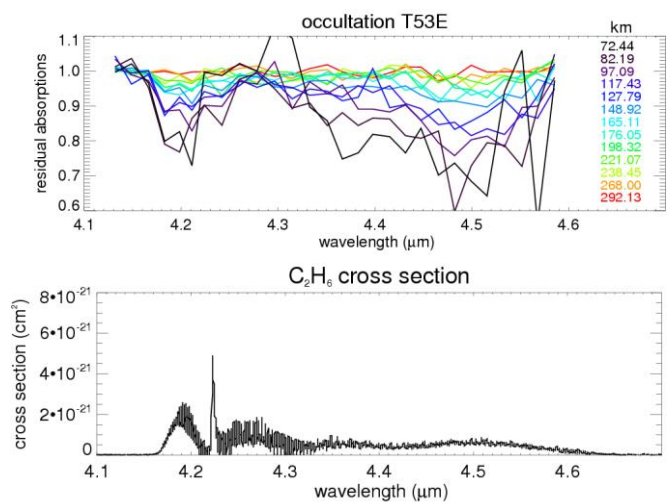

**Figure 28:** Same as in Fig. 27, for the 4.1–4.6 μm region. This is the same occultation of Fig. 23 (which shows the spectral fits of the same region). The two additional absorptions at 4.2 μm and between 4.3–4.5 μm are visible but the low signal-to-noise ratio does not allow us to follow clearly their shape and evolution with respect to altitude.

## 5 Discussion

In this paper we analyzed the full set of Titan solar occultations acquired by VIMS from the beginning of the Cassini mission. Several problems in the dataset reduced the usable occultations from 10 to 4. These, however, span different seasons and latitudes of Titan. We focused our study on the gaseous absorptions, extracting the vertical profiles of $CH_4$ and CO and looking at the residual absorptions in order to understand their origin. It must be remarked that the occultation dataset has also great potential to monitor the vertical distribution and seasonal evolution of the haze (see the last paragraph of Sect. 3.1 for a qualitative discussion), which has important consequences on the radiative budget of Titan. This will be the subject of a separate paper.

We have performed a careful analysis of the methane profiles extracted from different bands in the VIMS range. We find that the 1.4 and the 1.7 μm bands are much more reliable than the 2.3-μm used by Bellucci et al. (2009) in their study of the first solar occultation of VIMS. All occultations suggest a lower methane abundance than derived from GCMS/Huygens measurements. Our inferred methane abundance could be underestimated by the lack of forward-scattering in our modeled spectra. This is hinted by the bending of the methane profile towards higher abundances in the upper atmosphere, where the haze becomes more rarefied. However, a simulation of the effect of forward-scattering in transmission performed with a Monte-Carlo model shows that our results can underestimate the methane volume mixing ratio by at most 10%, while we obtain more than 20% difference between our values and the GCMS reference. Other recent works have measured the methane mole fraction in Titan's stratosphere. There is no general consensus. A reanalysis of CIRS data has found values as low as 1% (Lellouch et al. 2014), but we do not detect any latitudinal trend as they do. Measurements from the DISR/ULIS instrument find instead a methane abundance of $1.44^{+0.27}_{-0.11}$% between 29 and 135 km, in excellent agreement with GCMS (Bézard 2014). Our results are between 1.25% and 1.33% (after forward-scattering correction) and fall between these two extremes. Other measurements of the methane vertical profile and a more precise determination of methane's spectroscopic parameters at Titan's conditions are needed to understand the reasons for these discrepancies and to obtain reliable information on the stratospheric methane abundance.

CO is detected between ~50 and 170 km. We find, within the uncertainties, a uniform distribution of CO in the stratosphere and the lower atmosphere, with no latitudinal or seasonal variations. This result is in agreement with our knowledge of CO's behavior in Titan's atmosphere. We retrieve a global average abundance of 46±16 ppm, in good agreement with the stratospheric averaged CIRS value of 47±8 ppm (de Kok et al. 2007) and, within the uncertainties, with the VIMS limb measurements of 32±15 ppm (Baines et al. 2006).

We have confirmed the presence of an additional absorption within the 2 µm methane band, hinted at by Bellucci et al. (2009) but not identified, and we attributed it to gaseous ethane. This interpretation is supported by the excellent agreement with laboratory measurements of $C_2H_6$ absorption cross-section (Sharpe et al. 2004) throughout the whole VIMS spectral range. Sim et al. (2013) and Kim & Courtin (2013) attribute this band to hydrocarbon ices. However, they provide only a qualitative fit that does not reproduce completely their residuals in the 2 µm wavelength range. In addition, pressure and temperature conditions in Titan's atmosphere do not allow the formation and growth of hydrocarbon ices as high as 300 km, where we already detect some absorption.

The depths that we observe in VIMS spectra are compatible with the PNNL measurements. The PNNL mean cross sections around 2.3 and 2.42 µm are of the order of 3 to $4 \cdot 10^{-21}$ $cm^2$/molecule. At 180 km, assuming p = 1.25 mbar, T = 172 K and an ethane mole fraction of 12 ppmv (Vinatier et al. 2010a), we derive a $C_2H_6$ limb column density of $3.5 \cdot 10^{19}$ molecule $cm^{-2}$, which yields an absorption of 10–15 %, in good agreement with the residuals observed in occultation T78E at this altitude (Fig. 27). Unfortunately it is not possible to extract more precise information about these ethane bands, because PNNL spectra were recorded at pressures and temperatures higher than on Titan and also with smaller pathlengths. The PNNL cross sections, thus, cannot be directly extrapolated to Titan's environment. In any case, this study shows that the impact of ethane in Titan's infrared spectra has been seriously underestimated in previous studies.

We find that ethane affects other regions of Titan's near-infrared spectrum, through the presence of several bands here detected for the first time. Ethane cross sections measured by PNNL between 3.27–3.50 µm show a plateau of absorptions at $\sim 10^{-19}$ $cm^2$/molecule that are not included in the spectral databases but likely affect the 3 µm wavelength range in addition to all the other gaseous molecules and the C–H stretch of aliphatic hydrocarbons. A $10^{-19}$ $cm^2$/molecule cross section over the whole range 3.27–3.50 µm would essentially provide all the absorption seen in our residuals. However, we cannot really conclude that $C_2H_6$ alone is responsible for this missing absorption because of the limited spectral resolution of the PNNL spectra (0.11 $cm^{-1}$) and their different pressure (1 bar) and temperature (T > 278 K) conditions from Titan's ones. We note that $C_2H_6$ absorption could be responsible for the discrepancy between the depth of the residual absorption band and the haze density put into evidence by Rannou et al. (2010). The presence of a small ethane absorption band between 2.6–2.7 µm in the PNNL database supports the hypothesis that our residuals in this wavelength region, even if small, are a real feature. This is an important evidence because the wings of this band could affect part of the 2.7 µm surface window, which has proven difficult to model with our current knowledge of molecular absorption (Griffith et al. 2012, Hirtzig et al. 2013). Qualitative computations using a constant cross-section of $10^{-21}$ cm$^2$/molecule between 2.6–2.7 µm estimate an ethane band depth of 1–2% in nadir geometry. An absorption of this intensity may have a non-negligible impact on the shape of the 2.7 µm window. There is also a weak ethane band centered exactly on the 2 µm window, but it is unsure whether such a weak band could have an effect if observed in nadir geometry. Finally, we find that also the 1.7 µm methane band includes an additional ethane absorption at its center. This could lead to an overestimation of the methane abundance in our inversion of the 1.7 µm band. However, tests performing the inversion without the wavelengths affected by this additional absorption show that the effect is within the uncertainties.

The possibility of the presence of $C_3H_8$ absorptions must also be mentioned. Propane's abundance in Titan's atmosphere is approximately 1 ppm. Its spectrum has not been modeled in the near-IR, but PNNL measurements show that several bands of $C_3H_8$ exist in the VIMS range with cross-sections comparable to those of ethane (http://vpl.astro.washington.edu/spectra/c3h8pnnlimagesmicrons.htm). The shape of the $C_3H_8$ spectrum is however almost identical to $C_2H_6$, and it is impossible to disentangle the two molecules just by looking at VIMS' residual absorptions. Because of its lower abundance, in any case, the contribution of propane is significantly less than ethane.

The VIMS wavelength range is also rich in bands generated by stretching variations associated with aerosol particles. In addition to the features from aliphatic hydrocarbons, widely present around 3.4 µm, we tentatively attribute a very narrow and strong feature at 3.28 µm to PAHs. Further studies have to be carried out for a better characterization, but if confirmed, this would be the first detection of PAHs in Titan's stratosphere, following its identification in the upper atmosphere (Lopez-Puertas et al. 2013). This would have important consequences in constraining the processes of aerosol formation within the atmosphere of Titan. According to accretion modeling of aerosol formation on Titan, aromatic compounds in the thermosphere and upper atmosphere collide and aggregate to form bigger and more complex molecules (Lavvas et al. 2011). This process would decrease the PAHs' density in the lower atmosphere. A significant presence of PAHs in the stratosphere could suggest that aerosol coagulation is less effective than predicted by the models. Alternatively, the 3.28 μm band could be caused by aromatic molecules deposited at the surface of these aerosol particles.

We find also two significant additional absorptions at 4.2 μm and 4.3–4.5 μm. Their identification is less robust than for the above-described bands. The 4.2 μm feature seems not to increase its depth below ~90 km, so it could be generated by a gaseous species condensing around that altitude. Ethane possibly contributes to this absorption, but its band is too weak to be the only responsible for the 4.2 μm feature. In addition, ethane condenses only around 70 km. This band seems not to increase its depth below ~90 km, so it could be generated by a gaseous species condensing around that altitude. The 4.3–4.5 μm range contains a "plateau" of weak ethane absorptions added to bands from stretching variations of alkynes and nitriles that could create a broad signature. More observations with a better signal-to-noise ratio are needed for a more robust identification.

The seasonal evolution of Titan's vertical atmosphere observed with VIMS occultations will be pursued with new measurements before the end of the Cassini mission in 2017. Thus we will have profiles from almost half a Titanian year probing different latitudes. This dataset is a valuable tool to study the vertical behavior of gases and haze on Titan. It will permit to determine their properties and the identification of all the previously undetected spectral signatures with more precision.

**Acknowledgements**

We thank P. Lavvas, D.F. Strobel and P. Rannou for their interesting and useful insights, and the two anonymous reviewers whose suggestions helped to improve significantly the content and the readability of the paper. We thank the Agence Nationale de la Recherche (ANR Project "APOSTIC" n°11BS56002, France) for their support.